\documentclass[aps,prd,preprint,tightening,showpacs]{revtex4-1}
\usepackage{epsf,epsfig,graphics,graphicx,amsmath}
\usepackage[dvipsnames,usenames]{color}
\usepackage{comment}
\usepackage{ amssymb }
\usepackage[normalem]{ulem}

\newcommand{\vx}{\ensuremath{\vec{x}}}

\newcommand{\be}{\begin{equation}}
\newcommand{\ee}{\end{equation}}
\newcommand{\bea}{\begin{eqnarray}}
\newcommand{\eea}{\end{eqnarray}}
\begin{document}

\title{Primordial perturbations from ultra-slow-roll single-field inflation with quantum loop effects}

\author{Shu-Lin Cheng$^1$}
\email{shlcheng@gate.sinica.edu.tw}
\author{Da-Shin Lee$^2$}
\email{dslee@gms.ndhu.edu.tw}
\author{Kin-Wang Ng$^{1,3}$}
\email{nkw@phys.sinica.edu.tw}
\affiliation{$^1$Institute of Physics, Academia Sinica, Taipei 11529, Taiwan, R.O.C.}
\affiliation{$^2$Department of Physics, National Dong Hwa University, Hualien 97401, Taiwan, R.O.C.}
\affiliation{$^3$Institute of Astronomy and Astrophysics,
Academia Sinica, Taipei 11529, Taiwan, R.O.C. }

\date{\today}

\begin{abstract}
It is known that the single-field inflation with a transient ultra-slow-roll phase can produce a large curvature
perturbation at small scales for the formation of primordial black holes. In our previous work,
we have considered quantum loop corrections to the curvature perturbation and found that
the growth of these small-scale modes would affect the curvature perturbation at large scales
probed by cosmic microwave background observation.
In this work, we will further derive the constraints on the growing modes in the transition
between the slow-roll and the ultra-slow-roll phases under the effect of the loop corrections.
Our results would help clarify the recent controversy on whether or not the primordial-black-hole
formation from the single-field inflation is ruled out at one-loop level.
\end{abstract}

\maketitle
\newpage

\section{Introduction}
Inflation provides a solution to the horizon and flatness problems in our Universe and also furnishes a mechanism for generating scalar (curvature) quantum fluctuations that serve as the seeds of the primordial fluctuations in the matter density to form the rich network of cosmic structures we observe today. The inflationary paradigm that successfully explains the cosmological data relies
on the dynamics of a scalar field, the inflaton, evolving slowly during the inflationary stage
with the dynamics determined by a fairly flat potential. This simple inflationary scenario is referred to as slow-roll (SR) inflation. A distinct aspect of inflationary perturbations is that they are generated by quantum fluctuations of the scalar
field(s) that drive inflation.
 During  the stage of primordial
acceleration,  vacuum  fluctuations of the inflaton field
 are amplified to the cosmological scales, giving the approximately scale invariant spectrum of small density fluctuations on large scales, which are ultimately responsible for the CMB anisotropies and the large scale structure of our Universe~\cite{KT}.

The recent detection of gravitational waves from the merging of two 30 $M_{\odot}$ black holes~\cite{ABB}
has aroused the interest of primordial black holes (PBHs). This coalescence might be due to binary PBHs~\cite{bird,clesse,sasaki}. In addition, PBHs could comprise a considerable fraction of the dark matter, and thus leave imprints throughout the history of the Universe~\cite{CAR,MES,CAR1,Khlopov10} (see Ref.~\cite{SAS} for a review).
To generate sufficiently large perturbations in the power spectrum of the relevant short scales that later collapse into PBHs by gravitational pull, an inflation scenario must deviate from the SR regime. Ultra-slow-roll (USR) inflation was therefore proposed, invoking a transient USR phase in the single-field inflation~\cite{KIN,MAR,CHE,BYR}.
In our early works~\cite{SYU,CHE2}, we have considered the quantum loop effects to the primordial perturbations generated by a single scalar field during the dynamics of the SR-USR-SR inflation in the spatially flat gauge.
In Ref.~\cite{SYU}, we have studied the one-loop effects to the power spectrum in terms of the proper renormalized field quantities to cure the ultraviolent divergence by introducing the momentum cutoff. For a general potential model $V(\phi)$ of a scalar field $\phi$, the one-loop effects not only involve the first and second derivatives of the potential, $V'(\phi)$ and $V''(\phi)$, but also the third and fourth derivatives, $V'''(\phi)$ and $ V^{[4]} (\phi)$.
Our estimates have indicated significant one-loop corrections around the peak of the density power spectrum generated during the USR inflation. Subsequently, the work was extended to the Hartree resummation to account for the loop effects in a self-consistent manner that certainly is beyond the one-loop approximation~\cite{CHE2}, in which the momentum cutoff is set by the Hubble parameter to account for the sensitive infrared modes of superhorizon scales only in such a nearly de Sitter background~\cite{fordbunch}. Apart from involving $V'(\phi)$ and $V''(\phi)$, we have also included $V'''(\phi)$ and $V^{[4]}(\phi)$ to consider the loop effects to the background equation of motion for the inflaton mean-field and the mode equation of the field perturbation. Using specific inflation model potentials, we have shown that the presence of the USR phase may not only trigger a huge growth of the curvature perturbation at small scales that the seed the formation of PBHs, but also excessively produce
long-wavelength modes that leave the horizon in earlier times of the inflation, potentially at scales measured by cosmic microwave background (CMB) experiments. To avoid an overproduction of the perturbations at CMB scales,
it turns out that the amount of the produced small-scale perturbations may be too small for forming PBHs
to become a significant fraction of the dark matter.
Recently, there has been a debate on whether or not the PBH formation from the single-field USR
inflation is ruled out when the effect of one-loop corrections is taken into account,
with an open-ended conclusion that it may depend on the sharpness or smoothness of the transition
between the SR and USR phases~\cite{kri,rio,cho,cho2,kri2,rio2,cho3,fir,mot,cho4,fir2,fra,tas,fum}.
In this work, based on our early results in Ref.~\cite{CHE2}, we scrutiny the constraints on
the transition between the SR and USR phases and the duration of the USR phase, which help us find
the conditions for a successful inflation model that can accommodate both small- and large-scale
curvature perturbations for the potential production of PBHs and the CMB observation.

Our presentation is organized as follows. In the next section, we summarize the relevant equations
and the Hartree resummation scheme. The inflation model potentials are reexamined to find the
constraints on the USR phase in Sec.~\ref{sec3}. Concluding remarks and discussions are in Sec.~\ref{sec4}.

\section{Single-field inflation model and quantum loop effects}\label{sec2}
We briefly review the relevant equations from the single-field inflation model that include the quantum back reaction.
The details can be referred to Refs.~\cite{SYU,CHE2}.
The action under consideration is given by
\bea\label{action}
S=&& S_g+S_{\phi}\nonumber\\
=&&\frac{1}{2} \int d^4 x \sqrt{-g} R+\int d^4 x \sqrt{-g} \Bigg[ - \frac{1}{2} \;
\partial_{\mu} \phi \, \partial^{\mu}\phi  -V(\phi) \Bigg] \;,
\eea
with the unit of the reduced Planck mass $M_{Pl}^{-2}= 8 \pi G_N=1$.
In the course of inflation, the inflaton field can be split into the homogeneous mean field $\Phi_0$ and its quantum fluctuations $\varphi(\vec{x},t)$ as
\be\label{tad}
\phi(\vec{x},t)= \Phi_0(t)+\varphi(\vec{x},t)\;,
\ee
\noindent with
\be\label{exp}
\langle
\varphi(\vx,t)\rangle =0 \;.
\ee
The mean value is obtained with the non-equilibrium quantum state that later will
be specified to be the Bunch-Davies state often studied in the literature.
Here the further Hartree factorization approximation, namely $\varphi^3\rightarrow 3 \langle \varphi^2 \rangle \varphi$ and $\varphi^4\rightarrow 6 \langle \varphi^2 \rangle \varphi^2$,  is implemented~\cite{clu,kaz,SYU,CHE2}. The Hartree approximation certainly provides the resummation method in treating the contribution of $\langle \varphi^2 \rangle$ from the loop effects in a self-consistent manner and is therefore beyond the one-loop results~\cite{boyan3}. This approximation will become exact in the case of $N$ scalar field models as $N \rightarrow \infty$. Apart from the normal ultraviolent divergence in the momentum-integral that is cured by a regularization and a renormalization, in the case of minimally coupled massless inflaton fluctuations in de Sitter space-time,  $ \langle \varphi^2 \rangle $ also suffers from the infrared divergence~\cite{fordbunch} that may give sizable effects to the SR inflation \cite{boyan1} as well as the USR inflation~\cite{SYU}.
In this work, we will mainly focus on $\langle \varphi^2 \rangle$ that comes from the contributions of the infrared physical momentum modes $(k_{\rm phys} \leq H)$ while introducing a momentum cutoff $\Lambda=H$ in the momentum integration~\cite{CHE2}.

It is straightforward to generalize the equation of motion for the inflaton mean field including the loop corrections as
\be \label{phi_loop}
\ddot{\Phi}_0+3\,H \,\dot{\Phi}_0+V'(\Phi_0)+\frac{1}{2} V'''(\Phi_0) \, \langle \varphi^2 (\vec x,t) \rangle =0 \, ,
\ee
where the dot means the time derivative with respect to the cosmic time $t$.
The Friedmann equation just including $\langle \varphi^2 \rangle$ becomes
\bea \label{frieman_q}
H^2 &&=\frac{1}{3} \Big( \frac{1}{2} {\dot{\Phi}_0}^2  +V(\Phi_0)\Big)+\frac{1}{3} \bigg\langle \frac{1}{2} {\dot{\varphi}}^2 +\frac{\partial_i \varphi \partial_i \varphi}{2 a^2} +\frac{1}{2} V''(\Phi_0) \varphi^2 \bigg\rangle +\cdot \cdot \cdot \nonumber\\
&&\equiv H_0^2 +\delta H^2 \,.
\eea
Additionally, the spatial Fourier transform of the field
operator $\varphi_k $ obeys the mode equation, which can be obtained from the quadratic terms of $\varphi$ in the Lagrangian density~(\ref{action}) and the contributions from the metric perturbations in the spatially flat gauge at linear order together
with the further Hartree factorization approximation as \cite{bau}
\begin{equation}
\ddot{\varphi_k} + 3H \dot{\varphi_k} + \left[\frac{k^2}{a^2}+ V''(\Phi_0) - \frac{1}{a^3 M_{P}^2}\frac{d}{dt}\left( \frac{a^3\dot{\Phi_0}^2}{H} \right) {+\frac{1}{2} V^{[4]}(\Phi_0) \, \langle \varphi^2 \rangle }\right]\varphi_k = 0 .
\label{phieom2}
\end{equation}
In terms of $\zeta_k=\varphi_k/(\sqrt{2\epsilon})$, the mode equation becomes~\cite{clu,kaz,SYU,CHE2}
\begin{equation}
\ddot{\zeta_k} +\left( 3 + \eta \right) H \dot{\zeta_k} + \left[ \frac{k^2}{a^2} + \Delta_B \right]  \zeta_k = 0,
\label{MS2a}
\end{equation}
with the Hubble flow parameters  defined as
\begin{equation}\label{Hubble_flow}
\epsilon = - \frac{\dot{H}}{H^2},\quad \eta = \frac{\dot\epsilon}{\epsilon H}\, ,
\end{equation}
and
\begin{eqnarray}
\Delta_{B} &=&\Delta_{B3} + \Delta_{B4} \,,\quad {\rm with}\nonumber \\
\Delta_{B3} &=& V''(\Phi_0)-  \frac{1}{a^3}\frac{d}{dt}\left( \frac{a^3\dot{\Phi}_0^2}{H} \right) + \frac{H}{2}\dot{\eta} + \frac{H^2}{4}\eta^2 - \frac{H^2}{2} \epsilon \eta + \frac{3H^2}{2}\eta\,, \nonumber \\
\Delta_{B4}&=&{ V^{[4]}(\Phi_0) \, \langle \zeta^2 \rangle \, \epsilon }\,,
\label{Box1}
\end{eqnarray}
where $\Delta_{B}$ can be considered as an effective mass-squared term that can be negative.
The term $\Delta_{B3}$ originates from the cubic coupling that is included
as a back reaction to the equation of motion for $\Phi_0$ in Eq.~(\ref{phi_loop}).
The term $\Delta_{B4}$ which appears in the mode equation~(\ref{MS2a}) directly comes from the quartic coupling.
When ignoring the corrections from quantum fluctuations $\langle \varphi^2 \rangle$ or $\langle \zeta^2 \rangle$, we have $\Delta_{B3} \rightarrow 0$, recovering known Mukhanov-Sasaki equation~\cite{msvar}. After using both the mean field equation for $\Phi_0$~(\ref{phi_loop}) without the back reaction term and the Friedmann equation for $H_0$ in Eq.~(\ref{frieman_q}).
 $\Delta_{B4} \rightarrow 0$. We thus obtain the equation of motion for $\zeta$ in Eq.~(\ref{MS2a})
with $\Delta_{B}=0$. Nevertheless, including the quantum fluctuations will give nonzero $\Delta_B$ in Eq.~(\ref{Box1}).
This means that the $ \Delta_B$ is an indicator that could show the effects of the quantum fluctuations to the curvature perturbation, which we will discuss later.

For the choice of the standard initial Bunch-Davies vacuum states at early times $|k\tau| \gg 1$ with $\tau$ being the conformal time defined by $d\tau = dt / a$,
the mode functions behave the same as free-field modes in the Minkowski space-time.
For small $\epsilon$ and $\vert  \eta\vert $ with ignorable  $\langle \varphi^2 \rangle$, the Hubble parameter
$H$ is approximately a constant and thus the Universe is in an era of de Sitter space-time with the scale factor given by $a=e^{H_0t}=-1/(H_0\tau)$. As such, Eq.~(\ref{MS2a}) has a simple analytic solution,
\begin{equation}
\zeta_k= \frac{\pi^{1/2}H_0}{2k^{3/2}}\frac{1}{\sqrt{2\epsilon}}(-k\tau)^{3/2} H^{(1)}_{3/2}(-k\tau)\,.
\label{zetaana}
\end{equation}
In the slow-roll approximation, the power spectrum for the curvature perturbation is
\begin{eqnarray}
\Delta^2_{\zeta_k}&& \equiv  \frac{k^3}{2\pi^2} \left| \zeta_k \right|^2 \label{delta_exact}\\
&&=\frac{H^2}{8\pi^2 \epsilon} \, ,
\label{deltaappro}
\end{eqnarray}
where from Eq.~(\ref{Hubble_flow}) $\epsilon =\dot \Phi_0^2/2H_0^2$ by using the zero-order background equations.
In the next section, we will study how the back reaction of $\langle \varphi^2 \rangle $ affects the curvature  perturbation.

\section{Quantum fluctuations in SR-USR-SR inflation}\label{sec3}

In our previous work \cite{CHE2}, we consider the inflation models with an inflection point that will generate a large spike in the power spectrum as the source for the production of PBHs.
Two models are provided with
the Universe undergoing the SR-USR-SR inflation in which at a tree level
the created curvature perturbations are consistent with the Planck CMB measurements, namely the amplitude of the power spectrum of the curvature perturbation as well as the levels of the scalar spectral index, the tensor-to-scalar ratio, and the scalar spectral index running on large cosmological scales:
$\Delta_{\zeta_k}^2\simeq 2.3\times 10^{-9}$, $n_s\simeq 0.97$, $r<0.06$, and $|dn_s/d\ln k|<0.013$~\cite{planckinflation,bk18}, and also show a spike at relevant scales as the source for the PBH production.
However, when taking into account the loop effects,
we will find below that the two models exhibit different backreaction effects.
Although Model 1 can produce large enough density perturbations at relatively short-wavelength modes during the transient USR phase for seeding the PBH formation, the resulting loop effects give rise to an over-production of long-wavelength modes of relevance to CMB scales. In contrary, Model 2 can produce the long-wavelength perturbations consistent with observations as well as a sufficient amount of short-wavelength perturbations for forming PBHs. In this work, we will provide a detailed analysis in what conditions of the inflaton potential in particular with the back reaction included, one can have a successful SR-USR-SR transition that accommodates the CMB observations  at large scales, while producing small-scale perturbations relevant to the formation of PBHs.

 \subsection{Model 1}
\begin{figure}[htp]
\centering
\includegraphics[width=0.8\textwidth]{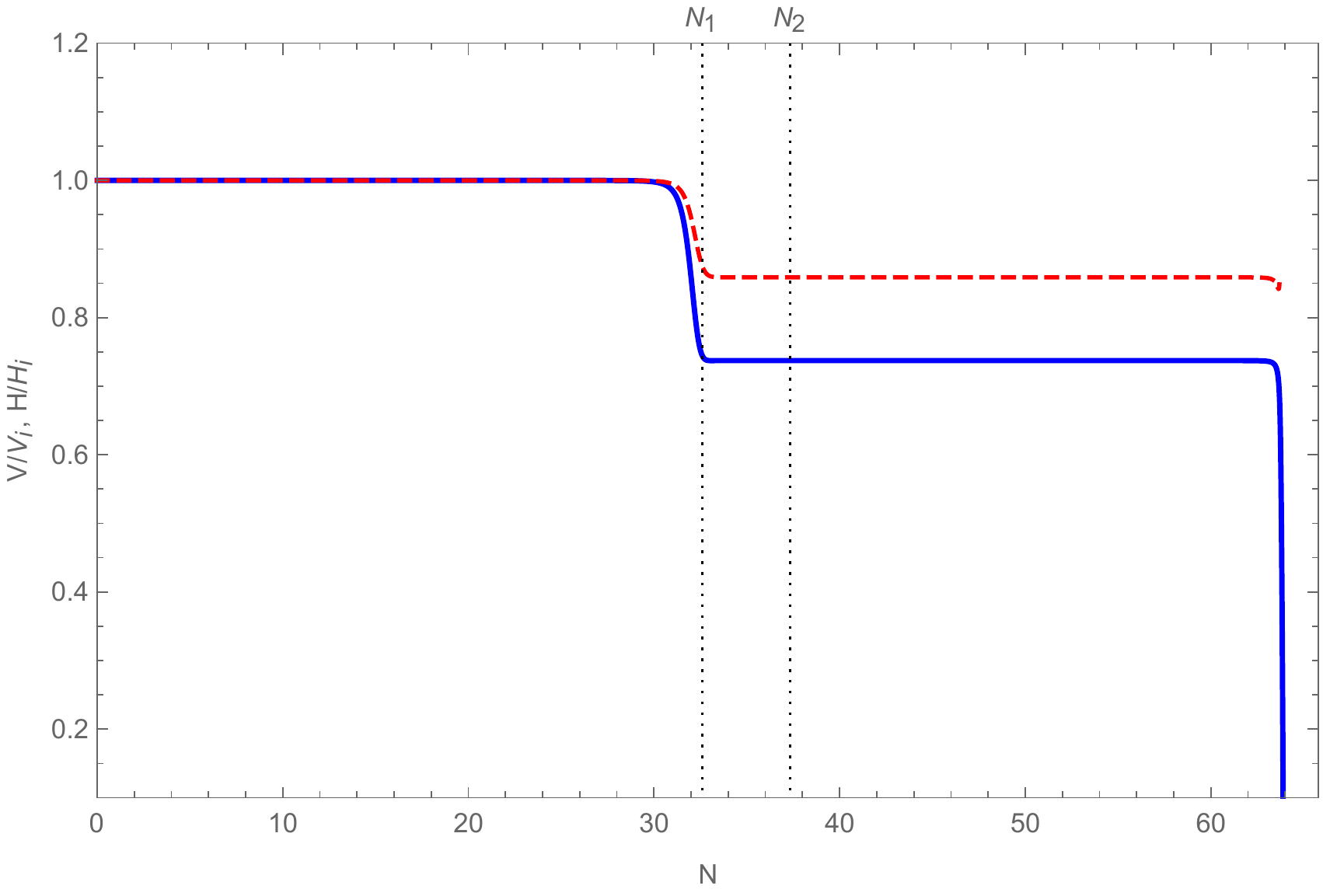}
\caption{Time evolution of the inflaton potential $V(\phi)/V_i$ in Eq.~(\ref{potential1}) (blue solid line) and Hubble parameter $H$ (red dashed line) with e-folding number $N$, where $V_i = V(\Phi_{0i})=1.6810 \times 10^{-14}$ and $H_i = 7.4855 \times 10^{-8}$.
$N_1$ and $N_2$ denote the beginning and end of the USR phase, respectively.
All dynamical variables in this figure and in the following figures are rescaled by the reduced Planck mass, $M_p=2.435\times 10^{18}$ GeV. }
\label{figpotential}
\end{figure}

The potential $V(\phi)$ of Model 1 is parametrized as~\cite{HER18}
\begin{equation}
V(\phi) = V_0 \left( 1+c_1 \frac{\phi}{c_\Lambda} + \frac{c_2}{2}\frac{\phi^2}{c_\Lambda^2} + \frac{c_3}{3!}\frac{\phi^3}{c_\Lambda^3} + \frac{c_4}{4!}\frac{\phi^4}{c_\Lambda^4} + \frac{c_5}{5!}\frac{\phi^5}{c_\Lambda^5} \right) ,
\label{potential1}
\end{equation}
with
\begin{eqnarray}
&&c_\Lambda = 0.3,\quad V_0 = {1.681} \times 10^{-14}, \nonumber \\
&&c_1 = {-0.7276} \times 10^{-4},\; c_2=0,\;c_3 = -0.52,\; c_4 = 1.0,\; c_5 = -0.6407.
\end{eqnarray}
The profile of the $V(\phi)$ is depicted in Fig.~\ref{figpotential}, where on the x-axis we have replaced the mean-field $\Phi_0 (N)$ by its corresponding e-folding number defined by $N=\int^t_0 H\left( t'\right) dt'$.

\begin{figure}[htp]
\centering
\includegraphics[width=0.8\textwidth]{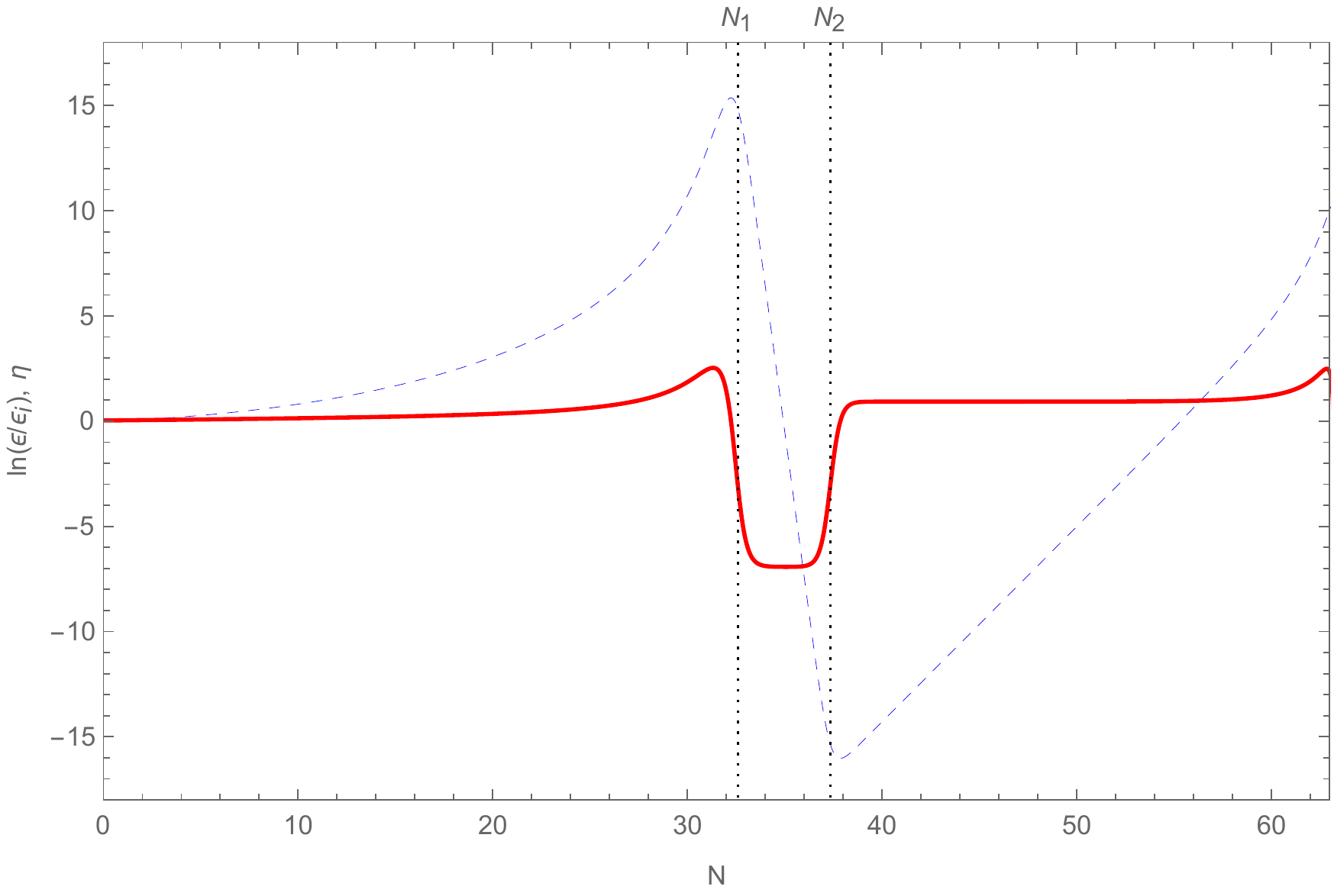}
\caption{Evolution of $\ln (\epsilon / \epsilon_i)$ (blue dashed line) and $\eta$ (red solid line) against e-folds $N$, where $\epsilon_i=3.08771\times 10^{-8}$ is the initial value of $\epsilon$ for Model 1.}
\label{figeps12}
\end{figure}

The initial conditions of the background field are set to be {$\Phi_{0 i} = 8.664 \times 10^{-4}$ and $ {\dot\Phi}_{0 i}= 1.860 \times 10^{-11}$} and those of the mode functions are  the  standard Bunch-Davies vacuum states.
Fig.~\ref{figeps12} shows the evolution of the Hubble flow parameters. The Universe starts from the SR inflation,
undergoes the USR inflation for a transient period from $N_1=32.6$ to $N_2=37.4$ during which $3+\eta<0$,
and then evolutes back to the SR  phase till the end of inflation at $N=63.8$.

\begin{figure}[htp]
\centering
\includegraphics[width=0.8\textwidth]{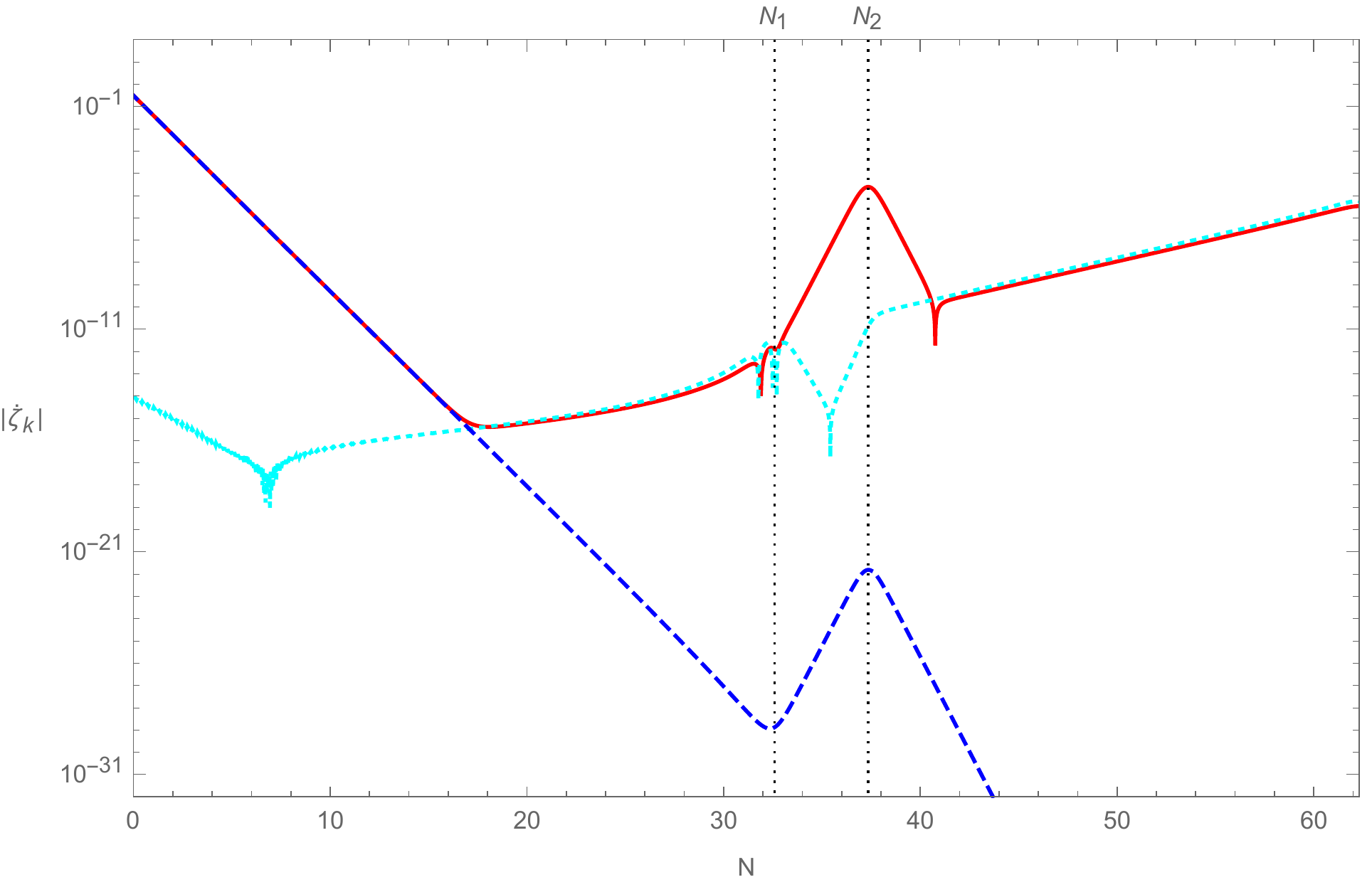}
\caption{Evolution of $| \dot{\zeta}_{k^*} |$ with loop corrections (red solid line), without loop corrections ($\Delta_{B}=0$, blue dashed line), and its approximated value from Eq.~(\ref{preB}) (cyan dotted line) against e-folds $N$, where $k^*/k_i=170.75$ and $k_i=H_i$. The value of $| \dot{\zeta}_{k^*} |$ is amplified during the USR interval between $N_1$ and $N_2$.}
\label{figdzeta}
\end{figure}

Fig.~\ref{figdzeta} shows the evolution of $|\dot{\zeta}_{k^*}|$ affected by the contribution of the loop correction 
$\Delta_B$ for a $k$-mode which crosses out the horizon at $N = 5$ ($k^* /k_i=170.75$).
Normally, the value of $|\zeta_k|$ should approach to a constant after the $k$-mode crosses out the horizon
($k < a H$) since $\ddot{\zeta_k}$ could be ignored in Eq.~(\ref{MS2a}) during the SR phase.
Meanwhile, the value of $|\dot{\zeta}_k|$ would be continuously decreasing during the inflation
because $|\dot{\zeta_k}| \sim \left| \zeta_k\right| k^2/(3Ha^2) $.
The effect of the loop correction appears at $N \sim 14$ for the $k^*$ mode, where $\Delta_B > (k/a)^2$.
The relatively large value of $\Delta_B$ before the USR phase, though suppressed by the slow-roll parameters, is resulted from a combination of the loop correction, the form of the potential, and the backreaction,
as depicted by Eq.~(\ref{Box3}) in Appendix~\ref{apx1}. Therefore, 
after $N \sim 14$ and before the USR phase, we can approximate the increase of $|\dot{\zeta_k}(N)|$ as
\begin{equation}\label{preB}
|\dot{\zeta_k}(N)| \sim \frac{|\Delta_B(N)|}{3H(N)}|\zeta_k(N)|\,.
\end{equation}
Apparently, this increase is significant compared to the case without higher-order corrections as seen in Fig.~\ref{figdzeta}.

During the USR phase ($3+\eta< 0$), the mode function can be approximated by $\ddot\zeta_k+ (3+\eta)H \dot \zeta_k\simeq 0$, so the $\dot{\zeta_k}$ could be amplified exponentially as
\begin{eqnarray}\label{fN}
&&\dot{\zeta_k}(N) \sim f(N)\,\dot{\zeta_k}(N_1)\,,\nonumber \\
&&f(N)= \exp\left[-\int_{N_1}^{N}(3+\eta) dN'\right]=
\exp\left[-\left(3N' + \ln\epsilon \right){\Big\vert}_{N_1}^{N}\right].
\end{eqnarray}
It was shown that the continuity of conformal weights across the transition
from the USR phase (with $\eta\simeq \eta_{\rm USR}$) to the second SR phase
(with $\eta\simeq \eta_{\rm 2SR}$) requires that
$\eta_{\rm USR}+\eta_{\rm 2SR}=-6$~\cite{wu2021}. In Fig.~\ref{figeps12},
it is evident that $\eta_{\rm USR}\simeq -7$ and $\eta_{\rm 2SR}\simeq 1$.
This implies that the $\dot{\zeta_k}$ reaches its maximum at $N=N_2$ and then
becomes exponentially damped shortly after $N=N_2$.
Therefore, for $N\gtrsim N_2$, we find from Eq.~(\ref{fN}) that
\begin{equation}\label{A}
{\zeta_k}(N) - {\zeta_k}(N_1)  = \dot{\zeta_k}(N_1) \int_{N_1}^{N} \frac{f(N')}{H(N')} dN'
\sim \frac{\dot{\zeta_k}(N_1)}{H(N_1)} A\,,
\end{equation}
where we have made $H(N')\sim H(N_1)$ and $A$ is an amplification factor estimated as
\begin{equation}\label{Afactor}
A= 2\int_{N_1}^{N_2} f(N')dN'\,.
\end{equation}
Normally, this amplification would not affect $|\zeta_k|$ for small $k$ modes because
the values of their $|\dot{\zeta_k}|$ are too small. So, they could still be ignored even after being amplified.
But the effect of the loop correction can provide these small $k$ modes with larger
values of $|\dot{\zeta}_k|$ at the beginning of the USR phase, which are given by Eq.~(\ref{preB}) as
\begin{equation}\label{B}
|\dot{\zeta_k}(N_1)| \sim \frac{|\Delta_B(N_1)|}{3H(N_1)}|\zeta_k(N_1)|\,.
\end{equation}
Hence, combining Eqs.~(\ref{A}) and (\ref{B}) gives
\begin{equation}
|{\zeta_k}(N) - {\zeta_k}(N_1)| \sim \frac{A|\Delta_B(N_1)|}{3H^2(N_1)}|\zeta_k(N_1)|\,.
\end{equation}
In order that the power spectrum of the curvature perturbation for small $k$ modes remains intact,
they have to meet the condition, $|{\zeta_k}(N) - {\zeta_k}(N_1)| < |\zeta_k(N_1)|$, which gives
\begin{equation}\label{con1}
A <  \frac{3 H^2(N_1)}{\vert \Delta_B(N_1)\vert }\,.
\end{equation}

Note that the above condition does not depend on the large scale.
The reason is the following. For any small $k$ modes, 
when $\Delta_B < (k/a)^2$, the time evolution of $\dot{\zeta_k}$ is given by 
$|\dot{\zeta_k}(N)| = |\zeta_k(N)| k^2/(3Ha^2)$. 
At horizon crossing (when $a=k/H$), the mode $\zeta_k$ gets frozen due to the Hubble friction, 
so from Eq.~(\ref{deltaappro}) the frozen value of $\zeta_k$ is given by 
$|\zeta_k|_c=(2\pi^2\Delta_{\zeta_k}^2)^{1\over2}/k^{3\over2}$, 
where $\Delta_{\zeta_k}^2\simeq 2.3\times 10^{-9}$ on large scales. 
At this moment, the amplitude of $\dot{\zeta_k}$ is $|\dot{\zeta_k}|_c=|\zeta_k|_c H/3$. 
Afterwards, we can rewrite the decrease of $\dot{\zeta_k}$ as $|\dot{\zeta_k}(N)| =|\dot{\zeta_k}|_c (k/aH)^2$. 
At the time when $\Delta_B = (k/a)^2$, $\dot{\zeta_k}$ has decreased to a value of 
$|\dot{\zeta_k}|_c \Delta_B/H^2$, which equals to $|\zeta_k|_c \Delta_B/(3H)$.
This value is the same as given by Eq.~(\ref{preB}). In other words, the presence of loop corrections
provides the superhorizon modes with an $k$ independent ratio between $|\dot{\zeta_k}|$ and $|\zeta_k|$
that equals to $\Delta_B/(3H)$ just before the USR phase.

\begin{figure}[htp]
\centering
\includegraphics[width=0.8\textwidth]{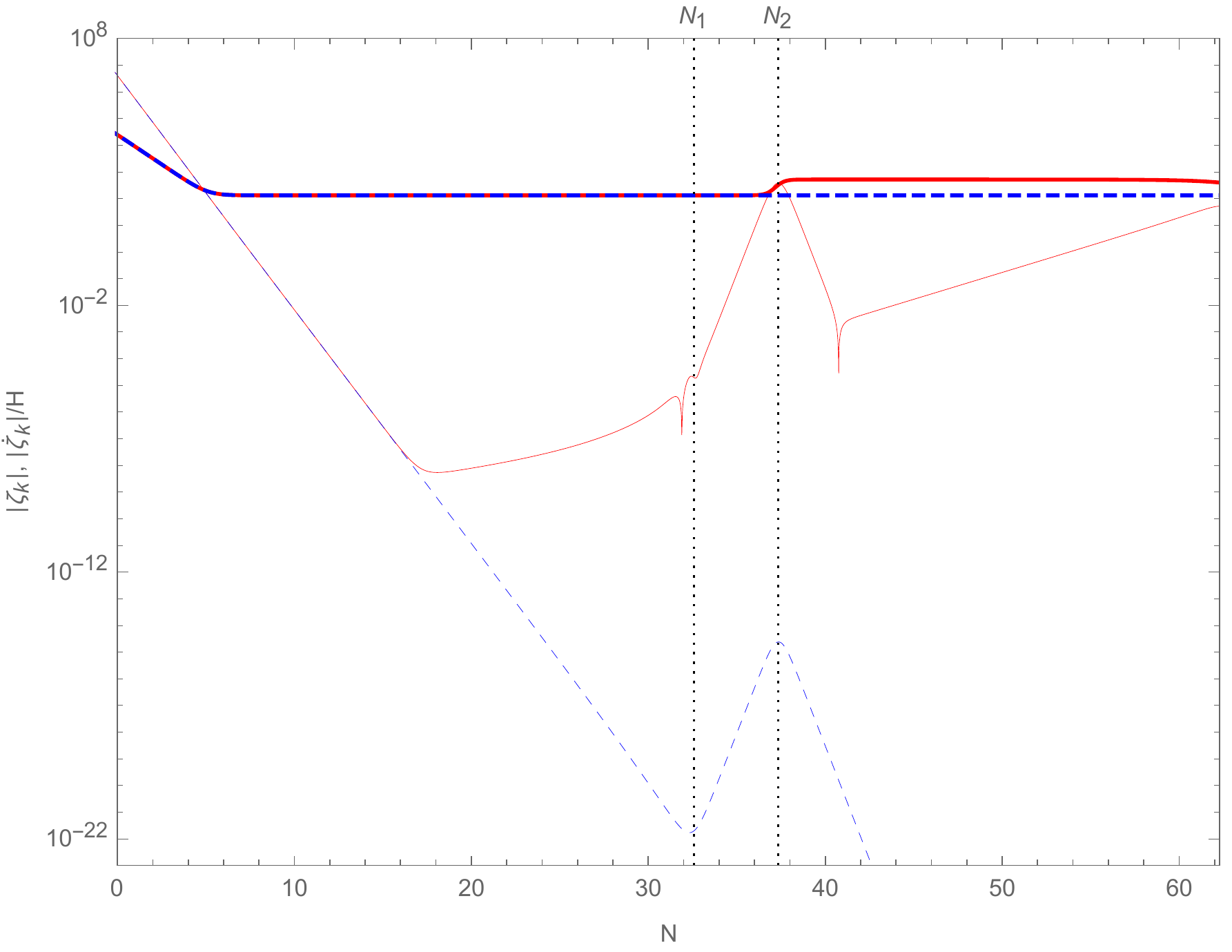}
\caption{Evolution of $\left| \zeta_{k^*} \right|$ and $| \dot{\zeta}_{k^*} |/H$ with loop corrections (denoted by the red solid line and the red thin solid line, respectively), and those without loop corrections (denoted by the blue dashed line and the blue thin dashed line, respectively) against e-folds $N$, where $k^*/k_i=170.75$ and $k_i=H_i$.
The $| \dot{\zeta}_{k^*} |$ with loop corrections is amplified to a value comparable with $H \left| \zeta_{k^*} \right|$ at $N\sim N_2$, resulting in a shift of the power spectrum for small $k$ modes.}
\label{figzeta}
\end{figure}

\begin{figure}[htp]
\centering
\includegraphics[width=0.8\textwidth]{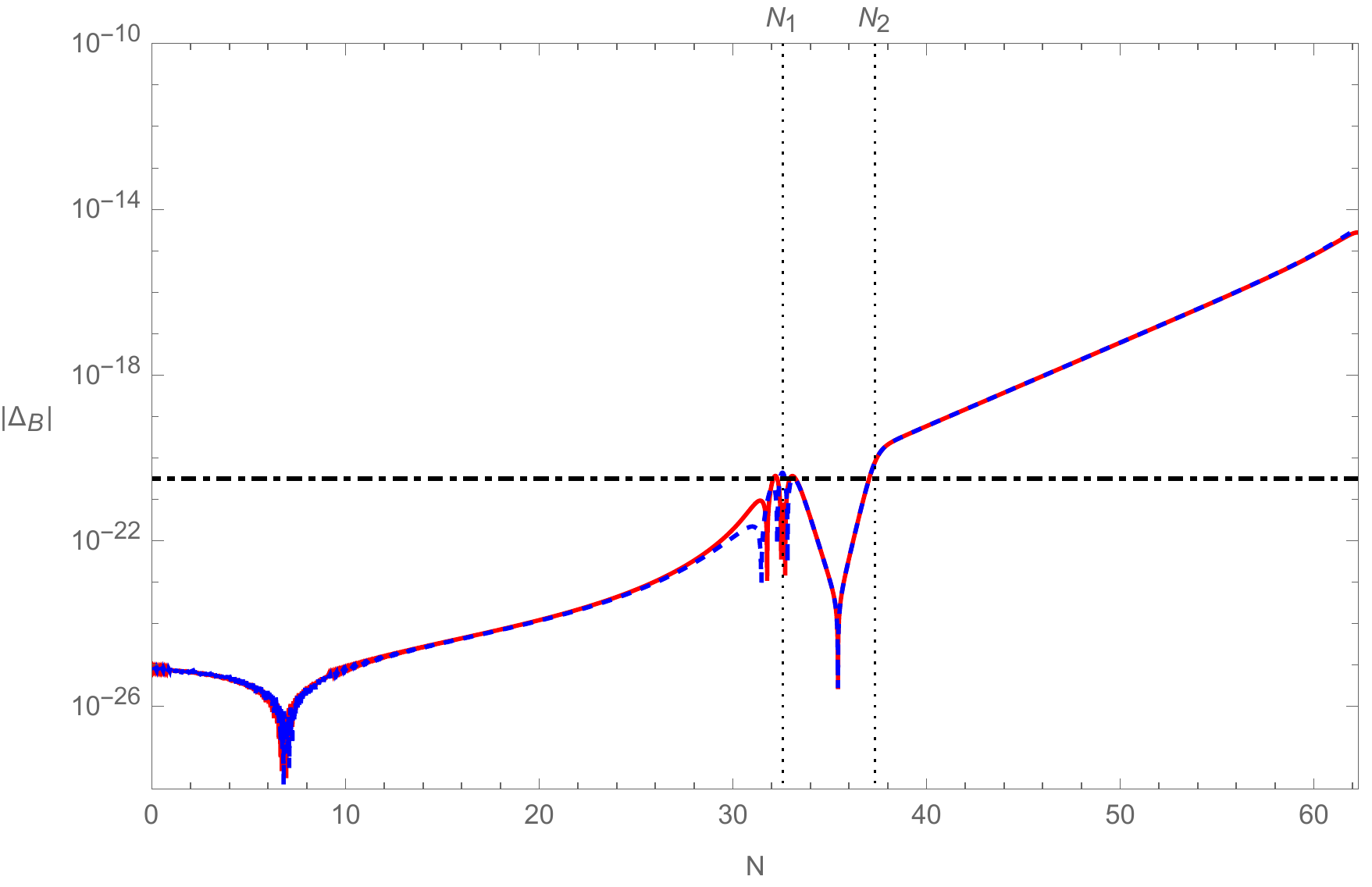}
\caption{Evolution of $\Delta_B$ (red solid line) and $\Delta_{B3}$ (blue dashed line) against e-folds $N$. Both are drawn with absolute values. The black dot-dashed line shows the constraint in Eq.~(\ref{con1}), $|\Delta_B| < 3H^2 A^{-1}$ at $N=N_1$, that the loop correction could affect the power spectrum for small $k$ modes. In Model 1, this constraint is violated, so it gives rise to a small-$k$ shift.}
\label{figDeltaB}
\end{figure}

The upper bound~(\ref{con1}) can be derived in a more intuitive way, though less accurate.
Fig.~\ref{figzeta} shows the evolution of $\left|\zeta_{k^*} \right|$ and $|\dot{\zeta}_{k^*} |/H$.
When the speed is larger than the displacement in a Hubble time,
i.e., $|\dot{\zeta_k}| > H\left| \zeta_k\right|$, $\left|\zeta_k\right|$ is increased.
To prevent the increment of $k$ modes at CMB scales, {$\left|\zeta_k\right|$ should not change much}
from $N_1$ to $N_2$. This means that at $N=N_2$ the maximum value of $|\dot{\zeta_k}|$ should be smaller
than $H\left| \zeta_k\right|$. Altogether, the following conditions must be satisfied:
\begin{equation}
|\zeta_k(N_1)| \sim |\zeta_k(N_2)|\,,\quad |\dot{\zeta_k}(N_2)| < H(N_2)\left| \zeta_k(N_2)\right|\,.
\end{equation}
The generical amplification behavior of $\dot {\zeta_k}$ from $N_1$ to $N_2$ during the USR inflation due to the large negative value of $\eta$ is encoded in Eq.~(\ref{fN}). We also consider
 the typical overdamped dynamics of the mode function $ {\zeta_k}$ just before $N_1$, the start of the USR inflation, giving the relation between $ \dot {\zeta_k} (N_1)$ and $  {\zeta_k} (N_1)$ in the period of the SR inflation in
 Eq.~(\ref{B}) for the mode $k <k_1$ (the mode $k_1$ leaving horizon at $N_1$)  where the loop effect becomes significant for these small $k$ modes.
Together with $H(N_1)\sim H(N_2)$, we find that
\begin{equation}
f(N_2) <  \frac{3 H^2(N_1)}{\vert \Delta_B(N_1)\vert }
\end{equation}
during the USR inflation from e-folding $N_1$ to $N_2$. Apparently the above heuristic derivations  can be applied to a general model, which derived constraint to  $ \Delta_B$ is treated in a model independent manner. Nevertheless, their precise values indeed depend on the model.
For Model 1, making the approximation that $\eta\simeq -7$ in Eq.~(\ref{fN}), we find that $A\simeq f(N_2)/2$.
Certainly, Model 1 violates the upper bound~(\ref{con1}), as shown in Fig.~\ref{figDeltaB}.

It is noted that $\Delta_B$ defined in Eq.~(\ref{Box1}) has no $k$-dependence, but presumably has $N$-dependence. 
As seen in Fig.~\ref{figDeltaB}, $\Delta_B$ remains a constant or increases before the USR phase. 
As such, the behaviour of $\dot\zeta$ is equally affected by the loop corrections for any small $k$ modes 
during the SR phase. The subsequent USR phase provides the extra exponential enhancement of $\dot\zeta$.
This enhancement of $\dot\zeta$ at large scales during the USR phase also has no
$k$-dependence. Therefore, the condition in Eq.~(\ref{con1}) for large-scale modes is scale independent.
We will see a similar behaviour of $\dot\zeta$ before the USR phase in Model~2 below.

Notice that the small-$k$ shift in the power spectrum depends on two factors.
The first is the duration and depth of the USR phase that determine the amplification factor
$A$ in Eq.~(\ref{Afactor}). However, the deepest $\eta_{\rm USR} \lesssim -6$ since after the USR phase
the Universe returns to the second SR inflation with $\eta_{\rm 2SR}\gtrsim 0$~\cite{CHE,wu2021}.
The second is the sharpness of the transition from the SR phase to the USR phase
that controls the value of $\Delta_B$ in Eq.~(\ref{B}), which is mostly contributed by
the $H\dot\eta/2$ term of $\Delta_{B3}$ in Eq.~(\ref{Box1}). The reader can refer to
Appendix~\ref{apx1} for a detailed analytic calculation of $\Delta_{B}$ in terms of
the Hubble flow parameters and the inflation potential, which allows us to evaluate
approximately $\Delta_{B}$ from the zero-order quantities.  { In Appendix~\ref{apx2}, we give a detailed description of the general behavior of the evolution of the curvature perturbation for each $k$ mode, using the dynamical equation for the mode function of relevance to each step of the SR-USR-SR inflation, with a particular attention to the typical feature of the occurrence of dip and peak. We show that the locations of dip and peak as well as the slopes connecting them depend on the model under consideration.}

\begin{figure}[htp]
\centering
\includegraphics[width=0.8\textwidth]{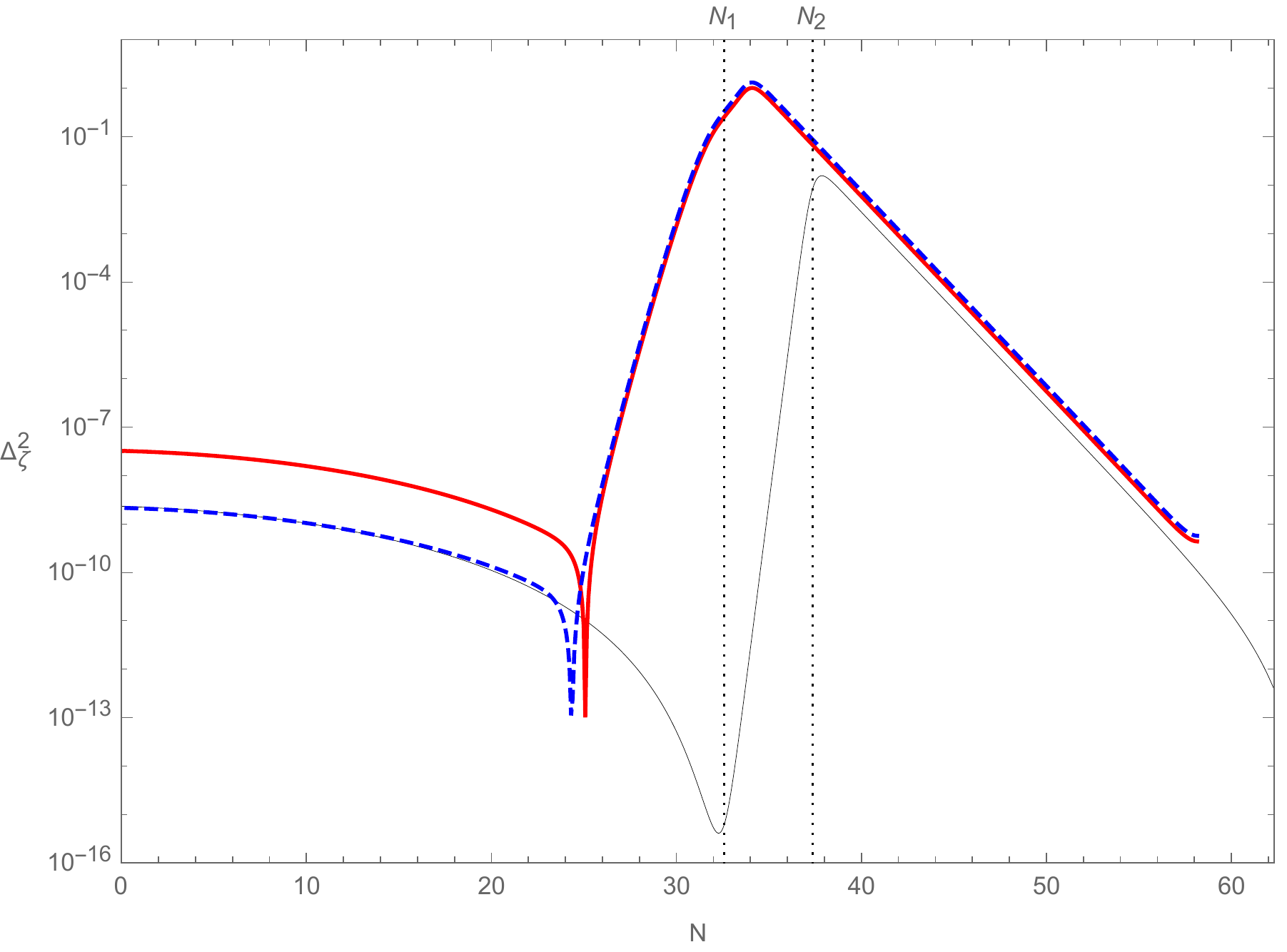}
\caption{Power spectrum of the curvature perturbation $\Delta_{\zeta_k}$ computed by Eq.(\ref{delta_exact}) with the solution of $\zeta_k$ without loop corrections (blue dashed line), with loop corrections (red solid line), and from the analytic result in Eq.~(\ref{deltaappro}) evaluated at the horizon-crossing time for each $k$-mode (black thin solid line), where $k=H e^N$.}
\label{figDeltaz}
\end{figure}

On the other hand, one needs a sufficiently large factor $A$ to boost the perturbations of
the short-wavelength modes for a significant production of PBHs. To obtain its lower bound,
let us now focus on the peak value of the power spectrum of the curvature perturbation.
In Fig. \ref{figDeltaz}, the power spectrum peaks at N=34.
The thin black line shows the approximated value of the power spectrum under the SR condition,
given by $\Delta_\epsilon^2 = H^2/(8\pi^2\epsilon)$ in Eq.~(\ref{deltaappro}),
which overlaps with the power spectrum of the corresponding $k$ mode at horizon crossing.
It could be enhanced by the growth of superhorizon modes during the USR phase,
leading to a higher peak value of the power spectrum~\cite{leach2001,CHE}.
The power spectrum is amplified to about $A^2$ times (since $\Delta^2_{\zeta_k}\propto | \zeta_k |^2 $) from the value of $\Delta_\epsilon^2$ for the $k$ modes that cross out the horizon near the beginning of the USR phase:
\begin{equation}
\Delta_{\zeta(peak)}^2 \sim A^2\Delta_{\epsilon}^2(N_1)\,,\quad{\rm with}\quad
\Delta_{\epsilon}^2(N_1)=\frac{H^2(N_1)}{8\pi^2\epsilon(N_1)}\,.
\end{equation}
This peak value should lie within the range,
\begin{equation}\label{con2}
\Delta^2_{\zeta(CDM)}<\Delta^2_{\zeta(peak)}<  \Delta^2_{\zeta(PBH)} ,
\end{equation}
where $\Delta^2_{\zeta(PBH)}$ is the upper limit derived from the astrophysical and
cosmological constraints on the abundance of PBHs (see, for example, Ref.~\cite{SAS}).
The $\Delta^2_{\zeta(CDM)}$ is supposed to be the lowest value of the power spectrum
with which the PBHs can make up a significant fraction of the cold dark matter.
Therefore, for a successful SR-USR-SR inflation model,
the factor $A$ has to satisfy both the constraints~(\ref{con1}) and~(\ref{con2}), which can be combined as
\begin{equation}
\frac{\Delta^2_{\zeta(CDM)}}{\Delta_{\epsilon}^2(N_1)} < A^2 <
{\rm Min}\left[\frac{\Delta^2_{\zeta(PBH)}}{\Delta_{\epsilon}^2(N_1)},  \frac{9 H^4(N_1)}{\Delta_B^2(N_1)}\right]\,.
\label{allcon}
\end{equation}

\subsection{Model 2}
\begin{figure}[htp]
\centering
\includegraphics[width=0.8\textwidth]{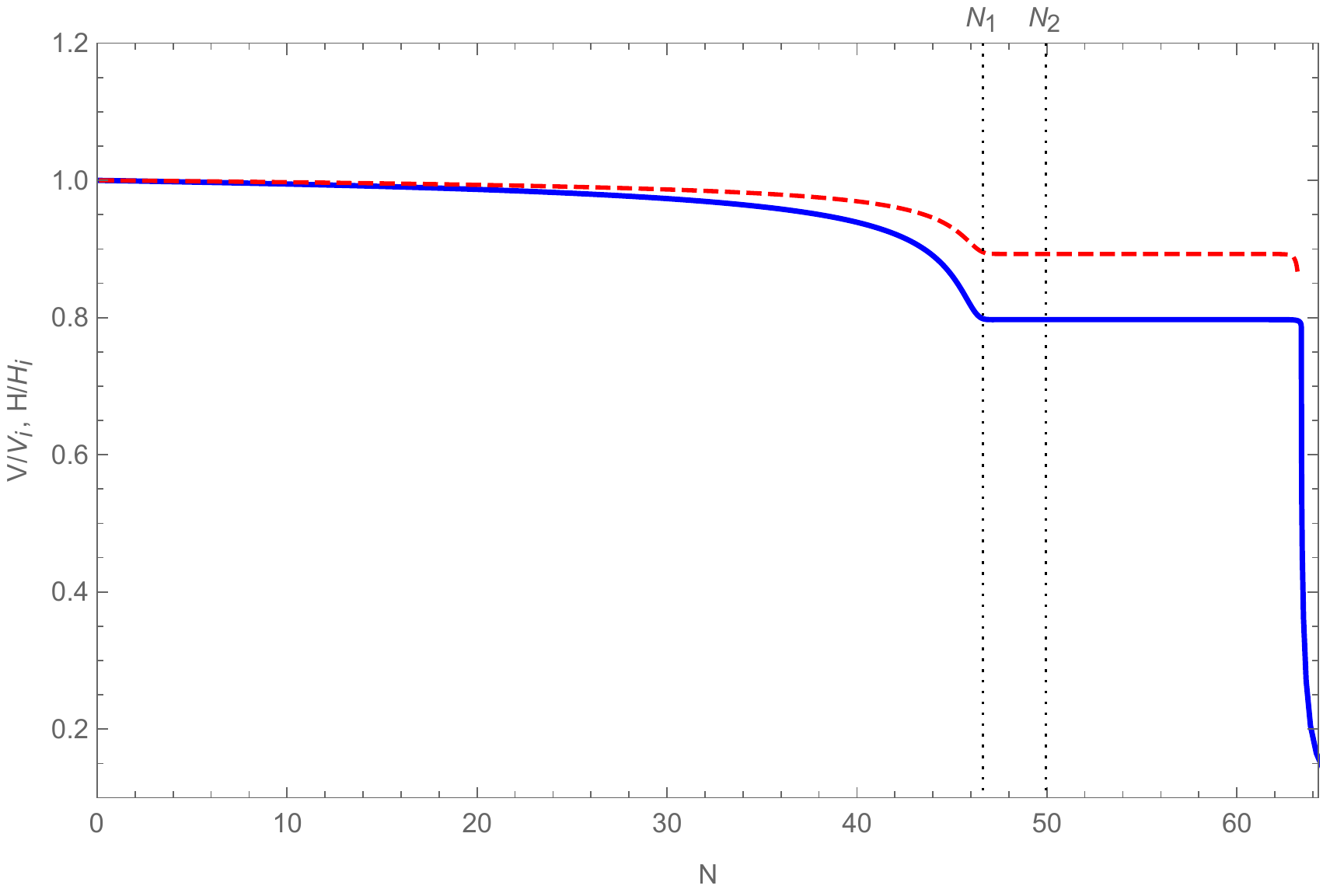}
\caption{Time evolution of the inflaton potential $V(\phi)/V_i$ in Eq.~(\ref{potential2}) (blue solid line) and Hubble parameter $H$ (red dashed line) with e-folding number $N$ for Model 2, where $V_i = V(\Phi_{0i})=1.2287 \times 10^{-10}$ and $H_i =6.34 \times 10^{-6}$.  All dynamical variables in this figure and in the following figures are rescaled by the reduced Planck mass, $M_p=2.435\times 10^{18}$ GeV. }
\label{figpotential_2}
\end{figure}

\begin{figure}[htp]
\centering
\includegraphics[width=0.8\textwidth]{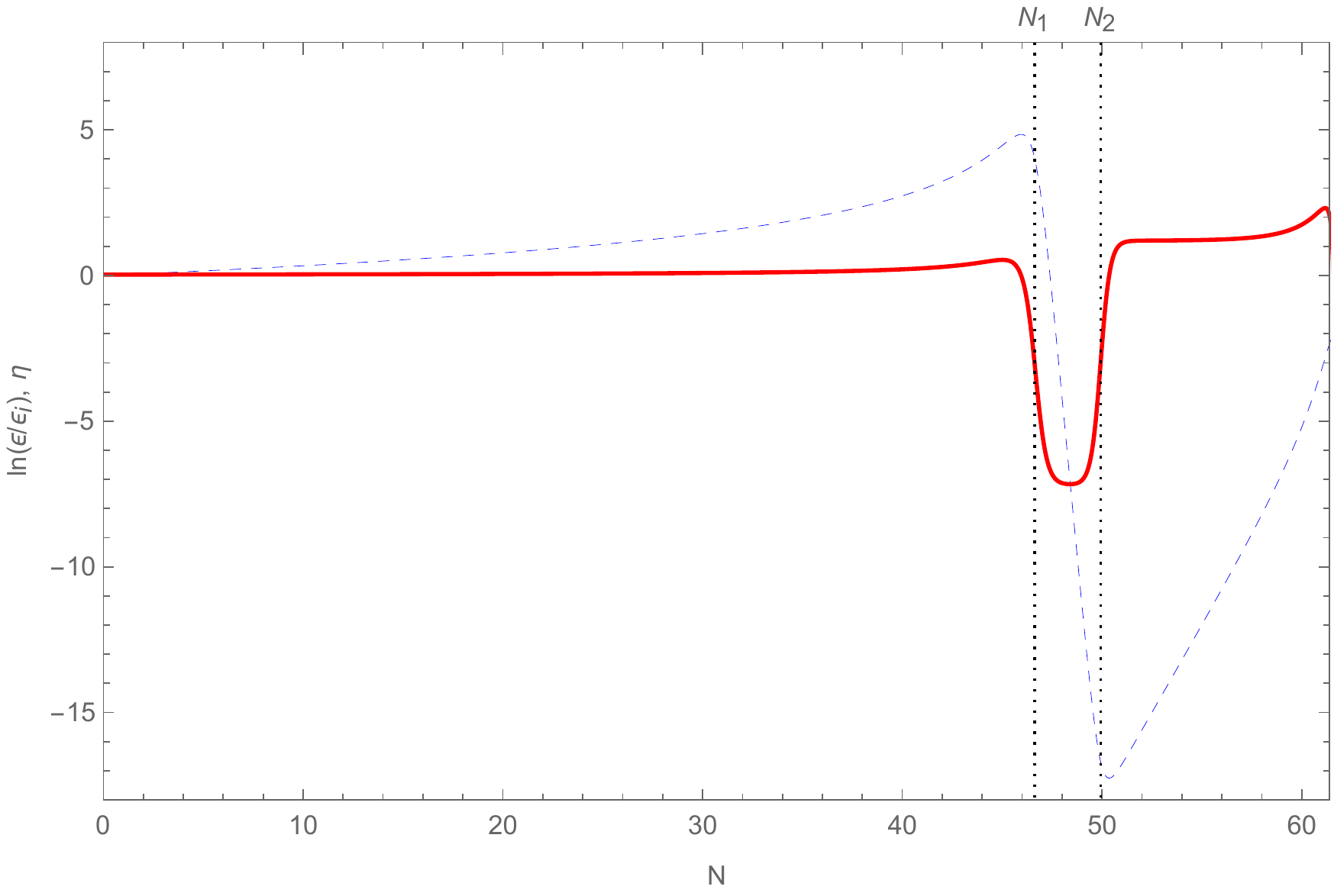}
\caption{Evolution of $\ln (\epsilon / \epsilon_i)$ (blue dashed line) and $\eta$ (red solid line) against e-folds $N$, where $\epsilon_i=2.25686\times 10^{-4}$ is the initial value of $\epsilon$ for Model 2.}
\label{figeps12_2}
\end{figure}

The second model we propose to study is parametrized as~\cite{juan17}
\begin{equation}
V(\phi) = \frac{\lambda}{12} \phi^2 \nu^2 \frac{6-4 a \frac{\phi}{\nu} + 3 \frac{\phi^2}{\nu^2}}{\left(1+b \frac{\phi^2}{\nu^2} \right)^2} ,
\label{potential2}
\end{equation}
where
\begin{eqnarray}
{\lambda = 1.12085 \times 10^{-6},\quad \nu=0.18, \quad a =0.7071, \quad b=1.5012\,,}
\end{eqnarray}
and the potential profile is shown in Fig.~\ref{figpotential_2}.
The initial conditions for the mean field are chosen to be $\Phi_{0i}=2.74 $ and $\dot\Phi_{0i}=-1.36\times10^{-7}$
and with the parameters above in the model the Hubble parameter $H_i=6.34\times 10^{-6}$.
The evolution of the Hubble flow parameters is plotted in
Fig.~\ref{figeps12_2}, which shows that the Universe undergoes the SR-USR-SR inflation.
The Universe starts from the SR inflation,
undergoes the USR inflation for a transient period from $N_1=46.6$ to $N_2=49.9$ during which $3+\eta<0$,
and then evolutes back to the SR  phase till the end of inflation at $N=62.3$.

\begin{figure}[htp]
\centering
\includegraphics[width=0.8\textwidth]{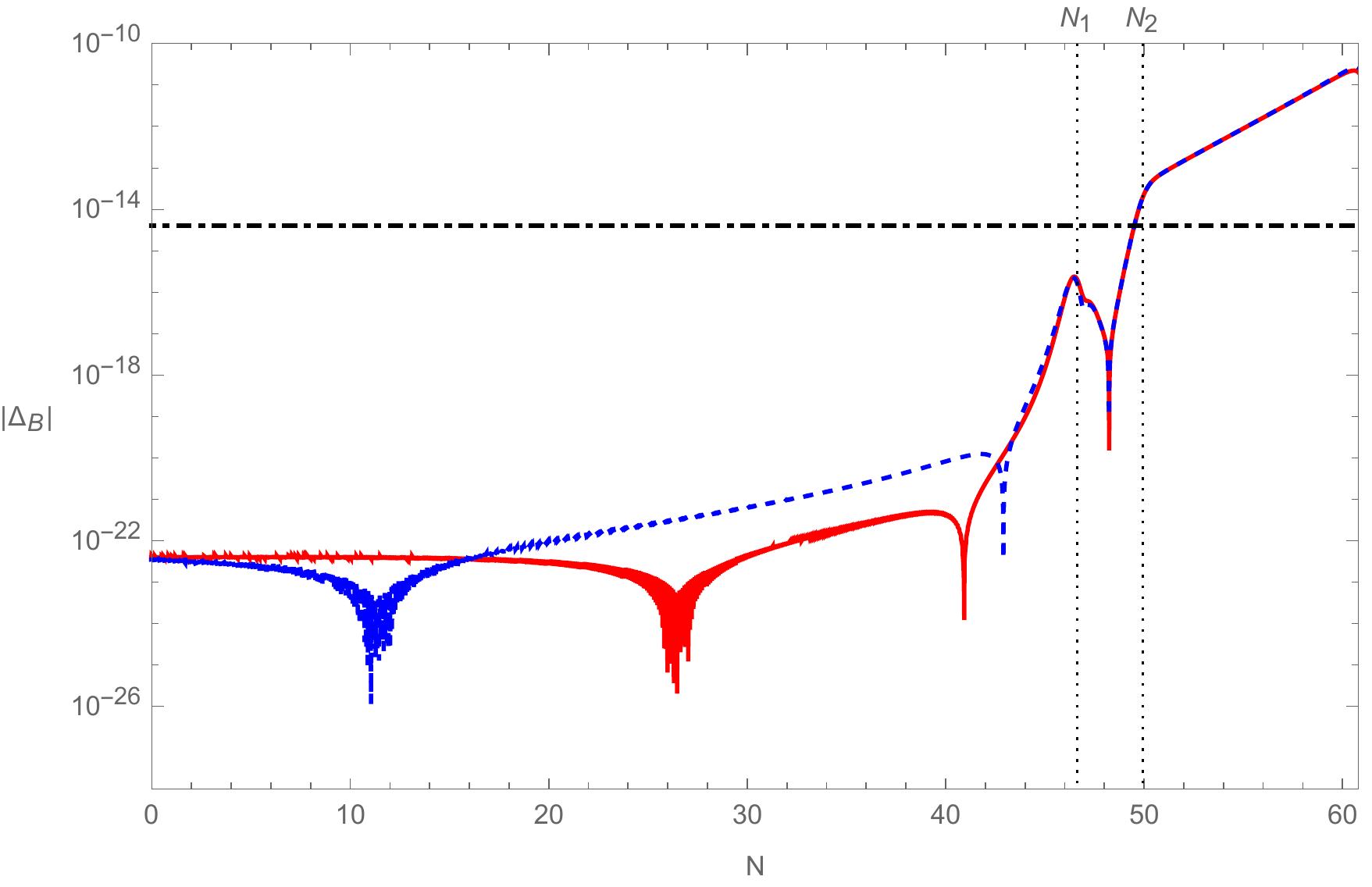}
\caption{Evolution of $\Delta_B$ (red solid line) and $\Delta_{B3}$ (blue dashed line) against e-folds $N$. Both are drawn with absolute values. The black dot-dashed line shows the constraint in Eq.~(\ref{con1}), $|\Delta_B| < 3H^2 A^{-1}$ at $N=N_1$, that the loop correction could affect the power spectrum for small $k$ modes. Model 2 satisfies this constraint.}
\label{figDeltaB_2}
\end{figure}

\begin{figure}[htp]
\centering
\includegraphics[width=0.8\textwidth]{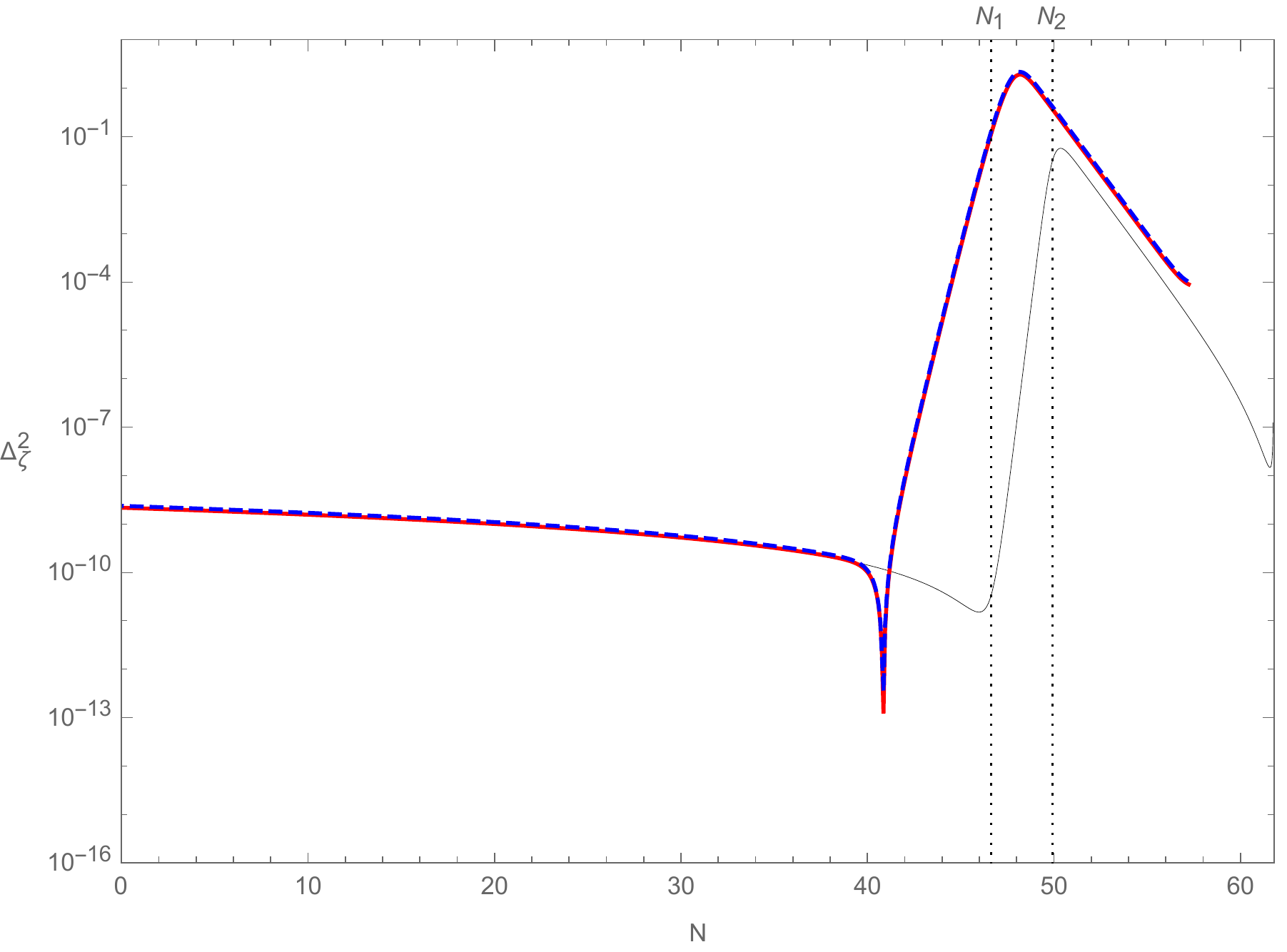}
\caption{Power spectrum of the curvature perturbation $\Delta_{\zeta_k}$ computed by Eq.(\ref{delta_exact}) with the solution of $\zeta_k$ without loop corrections (blue dashed line), with loop corrections (red solid line), and from the analytic result in Eq.~(\ref{deltaappro}) evaluated at the horizon-crossing time for each $k$-mode (black thin solid line) against e-folds $N$ for Model 2, where $k=H e^N$.
The dashed and solid curves overlap with each other.}
\label{figDeltaz_2}
\end{figure}

\begin{figure}[htp]
\centering
\includegraphics[width=0.8\textwidth]{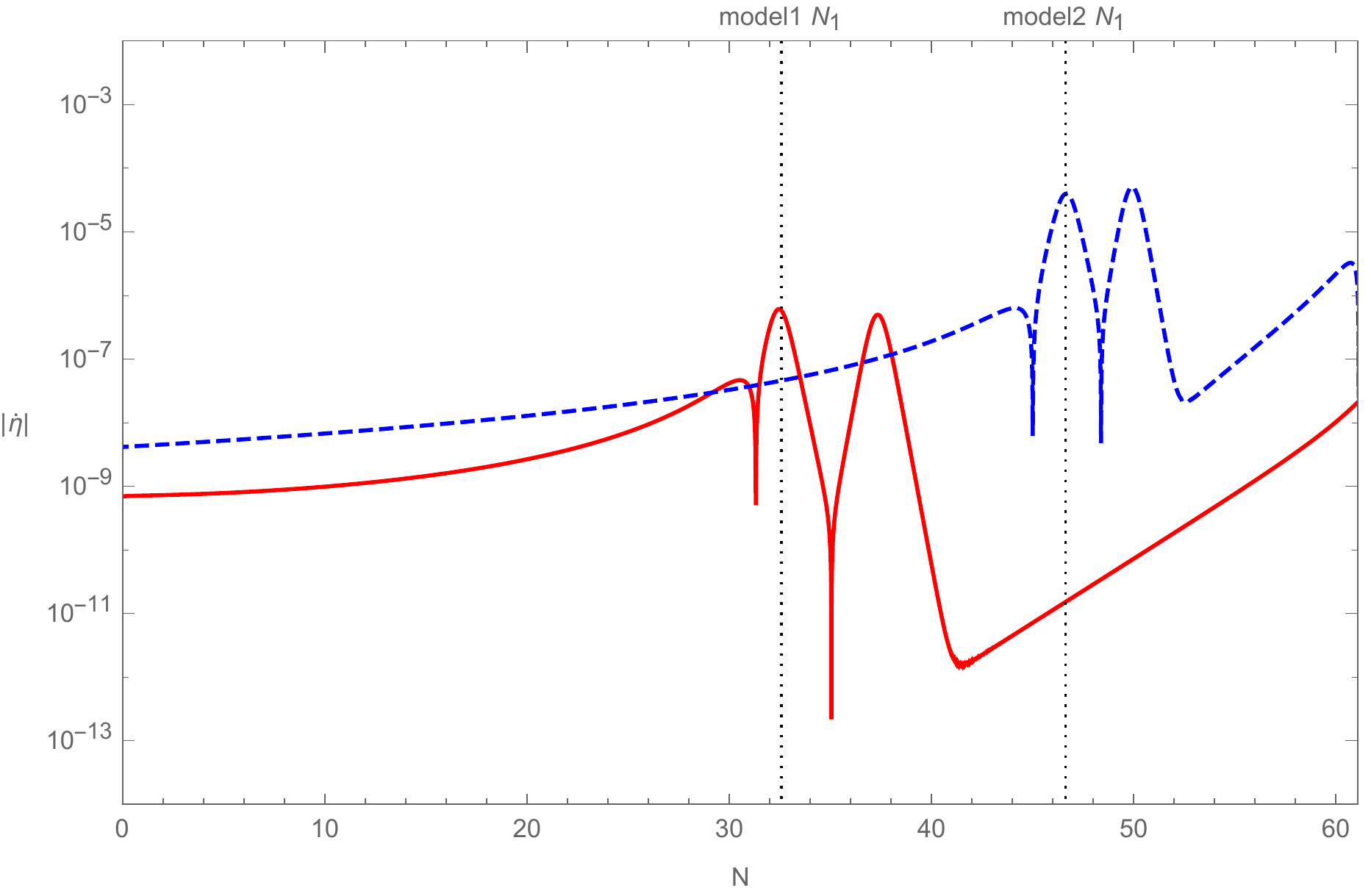}
\caption{Evolution of $\dot{\eta}$ against e-folds $N$ for Model 1 (red solid line) and Model 2 (blue dashed line).
Both are drawn with absolute values.}
\label{figdeps2}
\end{figure}

\begin{figure}[htp]
\centering
\includegraphics[width=0.8\textwidth]{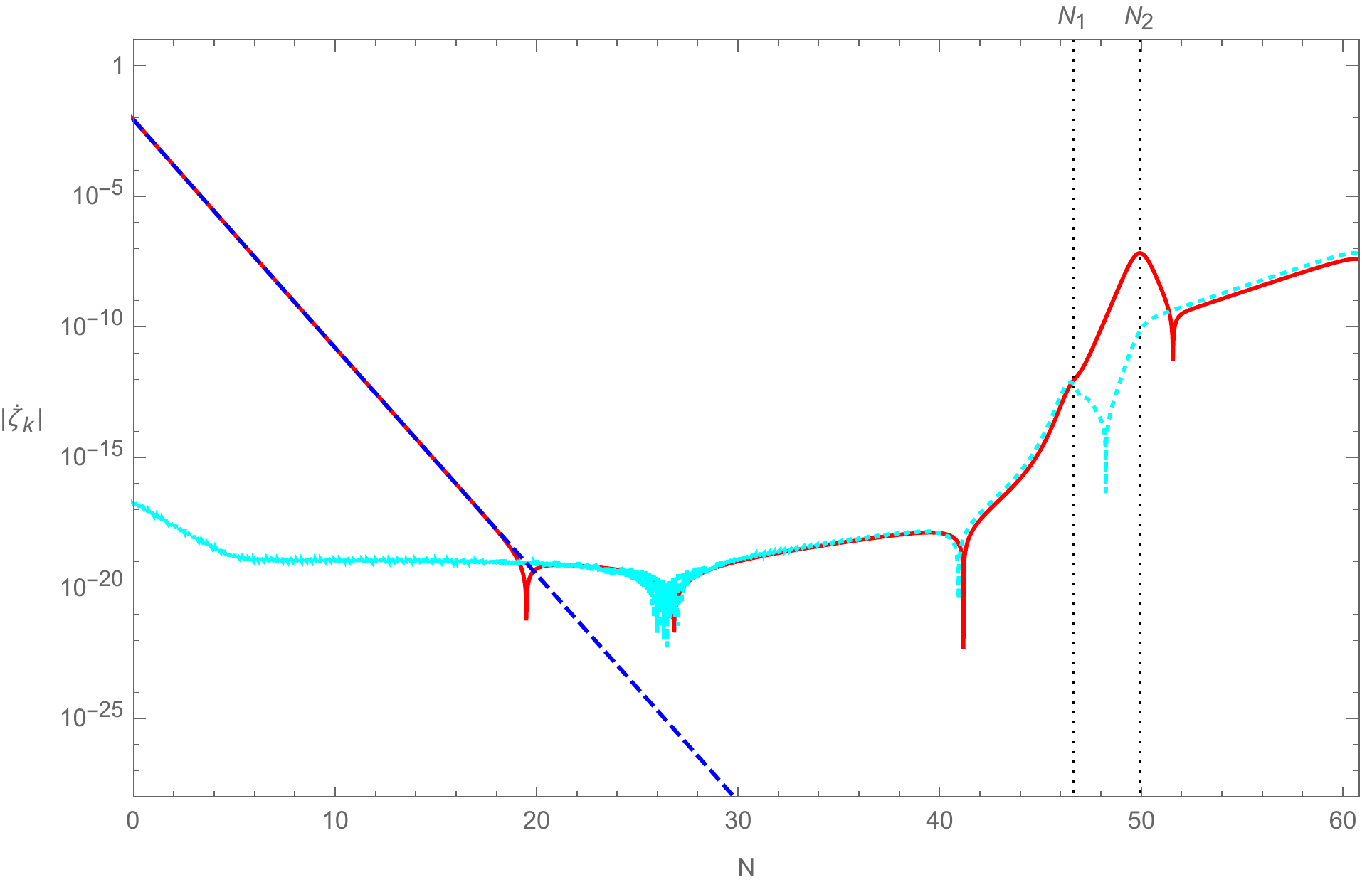}
\caption{Evolution of $| \dot{\zeta}_{k^*} |$ with loop corrections (red solid line), without loop corrections (blue dashed line), and its approximated value from Eq.~(\ref{preB}) (cyan dotted line) against e-folds $N$ for Model~2, where
 $k^*/k_i=151.19$ and $k_i=H_i$. The value of $| \dot{\zeta}_{k^*} |$ is amplified during the USR interval between $N_1$ and $N_2$.}
\label{figdzeta_2}
\end{figure}

\begin{figure}[htp]
\centering
\includegraphics[width=0.8\textwidth]{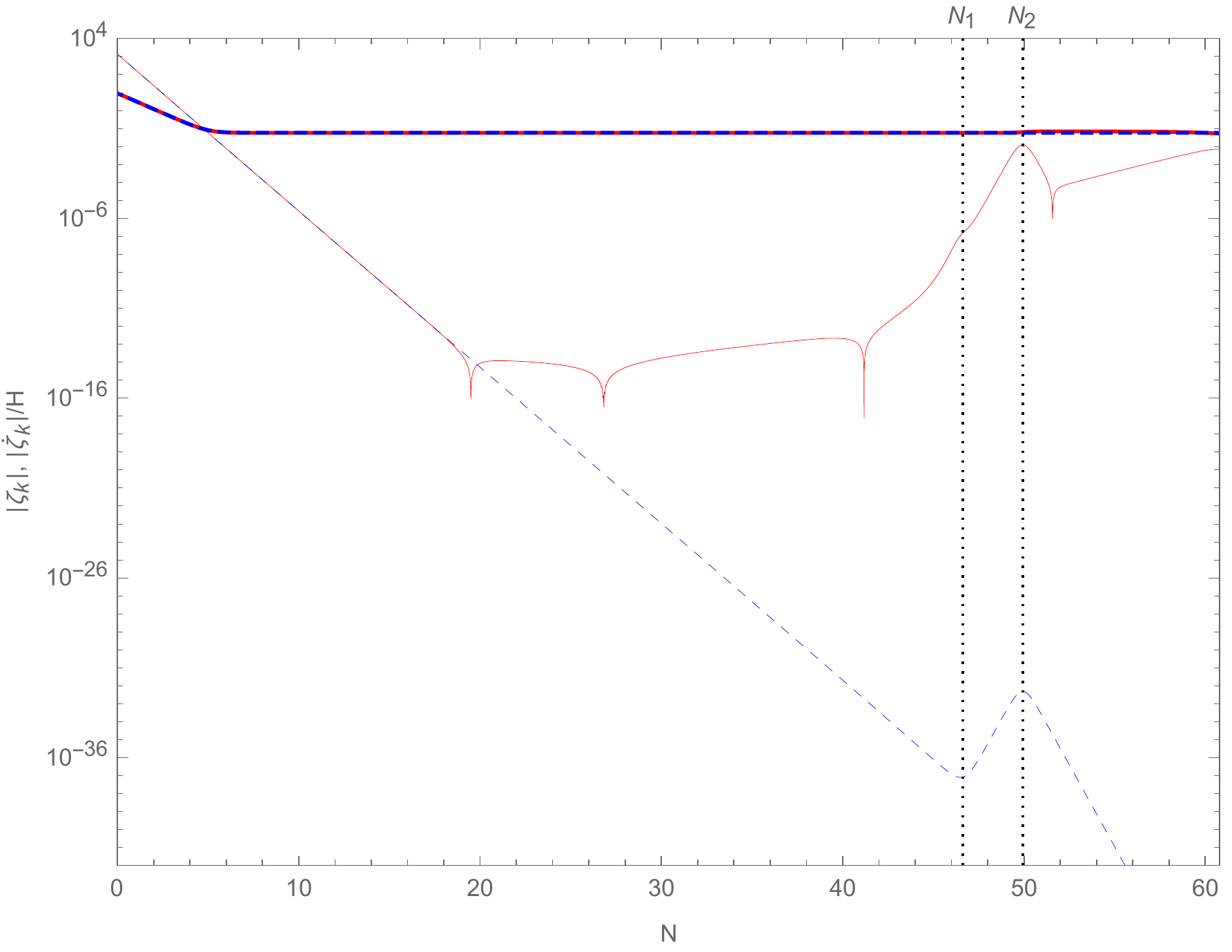}
\caption{Evolution of $\left| \zeta_{k^*} \right|$ and $| \dot{\zeta}_{k^*} |/H$ with loop corrections (denoted by the red solid line and the red thin solid line, respectively), and those without loop corrections (denoted by the blue dashed line and the blue thin dashed line, respectively) against e-folds $N$ for Model~2, where $k^*/k_i=151.19$ and $k_i=H_i$.
The $| \dot{\zeta}_{k^*} |$ with loop corrections is amplified but still below the value of $H \left| \zeta_{k^*} \right|$ at $N\sim N_2$, and therefore the loop correction will not affect the power spectrum.}
\label{figzeta_2}
\end{figure}

Then, one can compute the corresponding factor $A$ defined in Eq.~(\ref{Afactor}). We find that Model~2 obeys the constraint~(\ref{con1}) that prevents an overproduction of the large-scale perturbations as shown in
Fig.~\ref{figDeltaB_2}.
This is further verified by the power spectrum of the curvature perturbation in Fig.~\ref{figDeltaz_2},
which does not have a small-$k$ shift.
Notice that Model~2 does provide a sufficiently large amount of small-scale perturbations
for the formation of PBHs.
We find that the value of $\Delta_B$ is mainly determined by the difference
between the term $H\dot{\eta}/2$ with and without the loop correction,
so its value is ultimately related to the sharpness of $\eta$.
Fig.~\ref{figdeps2} shows that the value of $\dot{\eta}$ in the USR phase in Model~2 is higher than that in Model~1,
so Model~2 has a larger value of $|\Delta_B|$ in the USR phase than Model~1,
as seen in Fig.~\ref{figDeltaB} and Fig.~\ref{figDeltaB_2}.
However, the loop effect could be ignored in Model~2.
It is because the Hubble parameter $H$ has a greater value
and the amplification factor $A$ is not large enough in Model~2.
The values of $A$ are $8.34 \times 10^6$ and $4.89 \times 10^4$ for Model~1 and Model~2, respectively.
If the duration of the USR phase is longer, the amplification factor $A$ will become larger,
making the loop effect more manifest.
We have also drawn the time evolution of relevant quantities for Model~2
in Fig.~\ref{figdzeta_2} and Fig.~\ref{figzeta_2}.

\begin{figure}[htp]
\centering
\includegraphics[width=0.8\textwidth]{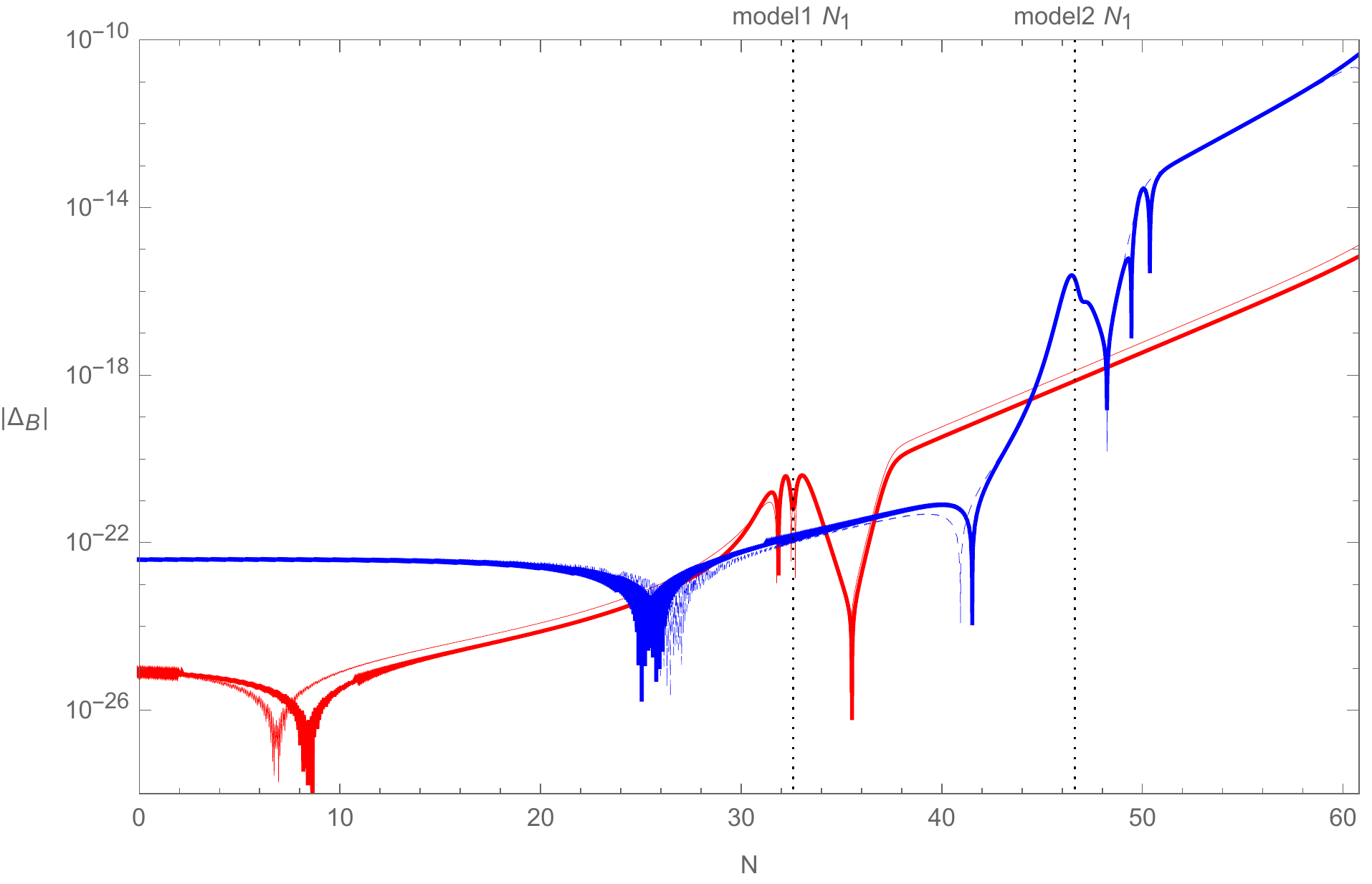}
\caption{Evolution of $\Delta_B$, drawn with absolute values, against e-folds $N$. In Model 1, the red thick and thin solid lines are from the approximated formula~(\ref{Box3}) and the full Eq.~(\ref{Box1}), respectively.
The corresponding blue thick and thin dashed lines are for Model 2.}
\label{figDeltaBapprox}
\end{figure}

\section{Conclusion}\label{sec4}
We have scrutinized the loop effects to the generation of the curvature perturbation in the single-field inflation with an ultra-slow-roll phase. We have improved and extended our previous numerical calculation~\cite{CHE}
that by virtue of the Hartree factorization,
self-consistently includes the cubic coupling as a back reaction to the motion of the inflaton mean field
and the quartic coupling as an effective mass term in the perturbation mode equation.
Our calculation is done in the spatially flat gauge that allows us to trace the time evolution of the inflaton perturbation under an inflation potential across the transition between the slow-roll and the ultra-slow-roll phases.
We could thus understand the requirements for a successful inflation model that can provide with small-scale
curvature perturbation for the production of primordial black holes while being consistent with the cosmic microwave background measurements on the primordial density perturbation. The condition in Eq.~(\ref{allcon}) summarizing all the requirements is the main result of the present work. It gives us the constraints on the sharpness across the transition from the slow-roll phase to the ultra-slow-roll phase as well as the duration of the ultra-slow-roll phase.
We have also found that the transition from the ultra-slow-roll phase to the final slow-roll phase,
being either a sharp or a smooth one, is irrelevant to the loop corrections. Although we derive the condition~(\ref{allcon}) using specific inflaton potentials, the underlying physics is transparent such
that it should apply to any ultra-slow-roll inflation models.

The loop correction to the generation of the curvature perturbation in the single-field inflation with an ultra-slow-roll phase has also been considered in Refs.~\cite{kri,rio,cho,cho2,kri2,rio2,cho3,fir,mot,cho4,fir2,fra,tas,fum},
in which they have mostly adopted the in-in formalism to compute the loop correction in the comoving gauge.
In principle, both methods of different choice of the gauges should be equivalent and thus the results should be comparable.
According to \cite{mad03}, the respective degrees of freedom for each of the gauges have linear relation under the gauge transformation (see Eq.(2.26) in \cite{mad03}). So, one can compare our approach in the spatially flat gauge with others, say in Ref.~\cite{kri}, in the comoving gauge in the sense of what loop diagrams are involved.
  Here we consider the cubic and quartic interactions and adopt the Hartree factorizations \cite{boyan3}.
In summary, the factorization of the cubic interaction gives the backreaction to the mean field equation of the inflaton.
The factorization of the quartic interaction presumably contributes to the one-loop diagrams to the power spectrum given by the so-called bubble diagram, which is local in time.  The adopted self-consistent approach  is to extend the one-loop bubble diagram to the daisy diagram. However, in Ref.~\cite{kri}
 the one loop diagrams resulting from the cubic interaction terms are given in terms of the second-order of the coupling constant, whose resulting loop contributions to the power spectrum  turn out to be nonlocal in time.
 Since two types of the diagrams give the same order-of-magnitude effects, one should consider  replacing the free field correlators in the computation of the one-loop diagrams in Ref.~\cite{kri} by the dressed ones given by the Hartree factorization to study their effects.
At last, in the condition~(\ref{allcon}), one needs to evaluate the effective mass of the perturbation mode at the beginning of the ultra-slow-roll phase. We have analytically derived an approximated formula for the time evolution
of the effective mass for a given inflaton potential that can be used to evaluate the effective mass from the zero-order quantities at any time during inflation. {The final words are that the achieved conditions  in Eq.~(\ref{allcon}) are based upon the dynamical equations for the mode functions relevant to the each step of the SR-USR-SR inflation but the detailed values  of the quantities depend on the models.}
This may give us a useful way to discriminate between ultra-slow-roll inflation models for the formation of primordial black holes.

\begin{acknowledgments}
This work was supported in part by the National Science and Technology Council (NSTC) of Taiwan, R.O.C.
under Grant Nos. 111-2112-M-259-006-MY3 (DSL) and 111-2112-M-001-065 (KWN).
\end{acknowledgments}

\appendix
\section{Differences induced by the loop correction}\label{apx1}
The loop correction induces small changes in the Hubble flow parameters:
\begin{equation}
\epsilon = - \frac{\dot{H}}{H^2} =\epsilon_0 + \delta\epsilon\,
\end{equation}
with
\begin{eqnarray}
\epsilon_0 &=& \frac{\dot{\Phi}_0^2}{2H^2} \label{eps0} \,,\\
\delta\epsilon &\simeq & \frac{\dot{\Phi}_0}{6H^3}V^{[3]}\left\langle \varphi^2 \right\rangle\,,
\label{epsdelta}
\end{eqnarray}
and
\begin{equation}
\eta =  \frac{\dot\epsilon}{\epsilon H} = \eta_0 + \delta\eta
\end{equation}
with
\begin{eqnarray}
\eta_0 &=& -6+\frac{\dot{\Phi}_0^2}{H^2}-\frac{\dot{\Phi}_0 V'}{H^3 \epsilon_0} \label{eta0} \,,\\
\delta\eta &\simeq & \frac{1}{6H^4 \epsilon}\left\lbrace \frac{1}{2}\left\langle \varphi^2 \right\rangle \left( V^{[3]}\left[ 3H\dot{\Phi}_0 \left( \epsilon_0 - 1 \right) - V' \right] + \dot{\Phi}_0^2 V^{[4]} \right) \right. \nonumber \\
&&\qquad\quad \left.  -V^{[3]^2}\left( \frac{1}{2} \left\langle \varphi^2 \right\rangle \right)^2 + \dot{\Phi}_0^2 V^{[4]} \left\langle \varphi\dot{\varphi} \right\rangle \right\rbrace \,,
\label{etadelta}
\end{eqnarray}
and its derivative
\begin{small}
\begin{eqnarray}
\dot{\eta}_0 &=& -12H\epsilon_0 - \frac{5\dot{\Phi}_0 V'}{H^2}+\frac{2\dot{\Phi}_0^2}{H}\epsilon_0 +\frac{\dot{\Phi}_0V'}{H^2 \epsilon_0}\left( 3+\eta_0  \right) + \frac{V'^2 - \dot{\Phi}_0^2 V''}{H^3 \epsilon_0} \,,\\
\delta\dot{\eta} &\simeq& \frac{1}{6H^4 \epsilon} \left\lbrace \frac{1}{2}\left\langle \varphi^2 \right\rangle \left( V^{[3]}\left[ -6\dot{\Phi}_0^3 +\frac{\dot{\Phi}_0^5}{H^2}-\frac{2V'^2}{\dot{\Phi}_0}+\frac{3\dot{\Phi}_0^2 V'}{H}+3H^2\dot{\Phi}_0 \left( 3\epsilon^2 - 6\epsilon+3+\eta \right) \right. \right. \right. \nonumber \\
&& \left.\left. \left. -\dot{\Phi}_0V'' + HV'\left( -7\epsilon+3+\eta \right) \right] + V^{[4]} \left[ H\dot{\Phi}_0^2 \left( 7\epsilon-9-\eta \right) - \dot{\Phi}_0 V'\right] + \dot{\Phi}_0^3 V^{[5]} \right) \right. \nonumber \\
&& \left. + \left( \frac{1}{2}\left\langle \varphi^2 \right\rangle \right)^2  \left( V^{[3]^2}\left[ -\frac{2V'}{\dot{\Phi}_0}+H\left( -7\epsilon+3+\eta \right)  \right] -4 \dot{\Phi}_0 V^{[3]}V^{[4]} \right) \right.  \nonumber \\
&& \left. + \left\langle \varphi \dot{\varphi} \right\rangle \left[ V^{[3]} H\dot{\Phi}_0 \left( 7\epsilon - 6 - \eta  \right) +2\dot{\Phi}_0^2 V^{[4]}\right] \right. \nonumber \\
&& \left. - 3V^{[3]^2} \frac{1}{2}\left\langle \varphi^2 \right\rangle \left\langle \varphi\dot{\varphi} \right\rangle +  \dot{\Phi}_0 V^{[3]} \left( \left\langle \dot{\varphi}^2 \right\rangle + \left\langle \varphi\ddot{\varphi} \right\rangle \right)\right\rbrace \, .
\label{dotetadelta}
\end{eqnarray}
\end{small}
The further expansion of $H$ can be done by writing $H=H_0+\delta H$ where we have found that $\delta H$ is ignorable in terms of the parameters we adopt.
These correction terms are so much smaller than the zero-order terms that they do not affect the inflation kinematics.
 In the mode equation of $\zeta_k$, since the sum of the zero-order terms in $\Delta_{B3}$ vanishes,
the loop correction terms in $\Delta_{B3}$  become relevant to the mode solutions.
After performing a lengthy but straightforward calculation,
we have found a fairly well approximation for $\Delta_{B}$,
which can be obtained from the zero-order mode solution $\zeta_k$ of Eq.~(\ref{MS2a})
with $\Delta_B=0$:
\begin{eqnarray}
\Delta_{B} &=& V''(\Phi_0)-  \frac{1}{a^3}\frac{d}{dt}\left( \frac{a^3\dot{\Phi}_0^2}{H} \right) + \frac{H}{2}\dot{\eta} + \frac{H^2}{4}\eta^2 - \frac{H^2}{2} \epsilon \eta + \frac{3H^2}{2}\eta + V^{[4]}\frac{1}{2}\left\langle \varphi^2 \right\rangle \nonumber \\
&\simeq& \frac{1}{12H^3 \epsilon_0} \left\lbrace \frac{1}{2}\left\langle \varphi^2 \right\rangle \left( V^{[3]}\left[  H^2\dot{\Phi}_0 \left( 10 \epsilon_0^2 - 12 \epsilon_0 + 18 + 3 \eta_0 \right) - \dot{\Phi}_0 V''  \right. \right.\right.  \nonumber \\
&& \qquad \qquad \qquad \qquad \qquad \left. + HV' \left( -6\epsilon_0 + 6 +\eta_0  \right)  \right] \nonumber \\
&& \qquad \qquad \qquad \quad \left. + V^{[4]} \left[ H\dot{\Phi}_0^2 \left( 8\epsilon_0- 6 -\eta_0 \right) - 3\dot{\Phi}_0 V'  \right] + \dot{\Phi}_0^3 V^{[5]} \right) \nonumber  \\
&&\quad\qquad \left. + \left\langle \varphi \dot{\varphi} \right\rangle \left( V^{[3]}\left[ H\dot{\Phi}_0 \left( 8\epsilon_0 - 9 - \eta_0   \right)  - 2V' \right] +2\dot{\Phi}_0^2 V^{[4]}\right) \right.  \nonumber  \\
&&\quad\qquad  \left. +  \dot{\Phi}_0 V^{[3]} \left( \left\langle \dot{\varphi}^2 \right\rangle + \left\langle \varphi\ddot{\varphi} \right\rangle \right)\right\rbrace \nonumber\\
&\simeq& \frac{1}{12H^3 \epsilon_0} \left\lbrace \epsilon_0\left\langle \zeta^2 \right\rangle \left( V^{[3]}\left[  H^2\dot{\Phi}_0 \left( 10 \epsilon_0^2 - 12 \epsilon_0 + 18 -6 \eta_0 + 7\epsilon_0\eta_0 \right) - \dot{\Phi}_0 V'' \right.\right.\right.  \nonumber \\
&& \qquad \qquad \qquad \qquad \qquad \left.  + HV' \left( -6\epsilon_0 +6 - \eta_0  \right)  +H \dot{\Phi}_0 \dot{\eta_0}  \right] \nonumber \\
&&   \qquad \qquad  \qquad\quad  \left. + V^{[4]} \left[ H\dot{\Phi}_0^2 \left( 8\epsilon_0- 6 +\eta_0 \right) - 3\dot{\Phi}_0 V'  \right] + \dot{\Phi}_0^3 V^{[5]} \right) \nonumber  \\
&&\quad\qquad \left. + 2\epsilon_0 \left\langle \zeta \dot{\zeta} \right\rangle \left( V^{[3]}\left[ H\dot{\Phi}_0 \left( 8\epsilon_0 - 9 +\eta_0   \right)  - 2V' \right] +2\dot{\Phi}_0^2 V^{[4]}\right) \right.  \nonumber  \\
&&\quad\qquad  \left. +  2 \dot{\Phi}_0 V^{[3]}  \epsilon_0 \left( \left\langle \dot{\zeta}^2 \right\rangle + \left\langle \zeta\ddot{\zeta} \right\rangle \right)\right\rbrace \, .
\label{Box2}
\end{eqnarray}
To obtain them,  we have ignored all high-order terms like $\left\langle \varphi^2 \right\rangle^2$, $ \left\langle \varphi^2\right\rangle \left\langle \varphi\dot{\varphi} \right\rangle$, and etc.
We can further omit all $\epsilon_0$ terms because $\epsilon_0 \ll1$, giving
\begin{eqnarray}
\Delta_{B} &\simeq& \frac{1}{6H\dot{\Phi}_0^2} \left\lbrace \frac{1}{2}\left\langle \varphi^2 \right\rangle \left( V^{[3]}\left[  3 H^2\dot{\Phi}_0 \left(  6 +  \eta_0 \right) - \dot{\Phi}_0 V'' + HV' \left( 6+\eta_0  \right)  \right] \right.\right.  \nonumber \\
&& \qquad \qquad \qquad \left. + V^{[4]} \left[ H\dot{\Phi}_0^2 \left( - 6 -\eta_0 \right) - 3\dot{\Phi}_0 V'  \right] + \dot{\Phi}_0^3 V^{[5]} \right) \nonumber  \\
&&\quad\qquad \left. + \left\langle \varphi \dot{\varphi} \right\rangle \left( V^{[3]}\left[ H\dot{\Phi}_0 \left(  - 9 - \eta_0   \right)  - 2V' \right] +2\dot{\Phi}_0^2 V^{[4]}\right) \right.  \nonumber  \\
&&\quad\qquad  \left. +  \dot{\Phi}_0 V^{[3]} \left( \left\langle \dot{\varphi}^2 \right\rangle + \left\langle \varphi\ddot{\varphi} \right\rangle \right)\right\rbrace \nonumber \\
&\simeq& \frac{1}{6H\dot{\Phi}_0^2} \left\lbrace \epsilon_0\left\langle \zeta^2 \right\rangle \left( V^{[3]}\left[ 6 H^2\dot{\Phi}_0 \left( 3 - \eta_0 \right) - \dot{\Phi}_0 V'' \right.\right.\right.  \nonumber \\
&& \qquad \qquad \qquad \qquad \quad \left.  + HV' \left(6 - \eta_0  \right)  +H \dot{\Phi}_0 \dot{\eta_0}  \right] \nonumber \\
&&   \qquad \qquad  \qquad  \left. + V^{[4]} \left[ H\dot{\Phi}_0^2 \left(- 6 +\eta_0 \right) - 3\dot{\Phi}_0 V'  \right] + \dot{\Phi}_0^3 V^{[5]} \right) \nonumber  \\
&&\quad\qquad \left. + 2\epsilon_0 \left\langle \zeta \dot{\zeta} \right\rangle \left( V^{[3]}\left[  H\dot{\Phi}_0 \left( -9+\eta_0 \right)  - 2V' \right] +2\dot{\Phi}_0^2 V^{[4]}\right) \right.  \nonumber  \\
&&\quad\qquad  \left. +  2 \dot{\Phi}_0 V^{[3]}  \epsilon_0 \left( \left\langle \dot{\zeta}^2 \right\rangle + \left\langle \zeta\ddot{\zeta} \right\rangle \right)\right\rbrace \, .
\label{Box3}
\end{eqnarray}
This is the final relatively simple form that can be used to evaluate $\Delta_{B}$
from the zero-order quantities. The results are shown in Fig.~\ref{figDeltaBapprox}.

\section{Dip, peak, and slopes of the power spectrum of curvature perturbations}\label{apx2}
The power spectrum of the curvature perturbation can be obtained  from the evolution of the $\zeta_k$.
The shape of the power spectrum has been largely discussed. For a summary, the reader may refer to Ref.~\cite{qin23} and references therein.
In this appendix, we plan to give more elaborated estimates of the locations of the dip and peak as well as the slopes of the power spectrum,
using the results obtained in the main text.

Apparently, both of the dip and peak start to appear during the USR regime whereas the flat spectrum is developed in the period of the first SR phase.
So, the dynamics of the modes with $k<k_1$ at  $N$ before the USR phase, where the mode $k_1$ leaves the horizon at $N_1$, is given by
\begin{eqnarray}
\left( 3+\eta \right)H \dot{\zeta_k} +  \left( \frac{k^2}{a^2} + \Delta_B \right) \zeta_k(t) \simeq 0 \, .
\end{eqnarray}
Thus,
\begin{equation}
\dot{\zeta_k} (N_1)  \simeq  -\frac{H(N_1)}{3} \left( \frac{k^2}{k_1^2} + \frac{\Delta_B (N_1)}{H^2 (N_1)} \right) \zeta_k(N_1)\,,
\end{equation}
where the value $\eta$ has been ignored in the SR phase. According to Eq.~(\ref{fN}),  $ \vert \dot{\zeta_k} (N)\vert$ grows by a factor of $A$ in the USR regime, giving $ \vert {\zeta_k} (N)\vert$ at $N_2$ as
\begin{eqnarray}
|\zeta_k(N_2)|^2 \simeq |\zeta_k(N_1)|^2 \left[ 1- \frac{A}{3} \left( \frac{k^2}{k_1^2} + \frac{\Delta_B(N_1)}{H^2(N_1)} \right) \right]^2 .
\end{eqnarray}
Then, the power spectrum at $N_2$ is obtained as
\begin{eqnarray}
 \Delta_{\zeta_k}^2(N_2) &=& \frac{k^3}{2\pi^2}\left\vert \zeta_k (N_2) \right\vert^2 \nonumber \\
&\simeq & \frac{k^3}{2\pi^2} |\zeta_k(N_1)|^2 \left[ 1- \frac{A}{3} \left( \frac{k^2}{k_1^2} + \frac{\Delta_B(N_1)}{H^2(N_1)} \right) \right]^2 \nonumber \\
&\simeq & \Delta_{\epsilon}^2(N_k) \left[ 1- \frac{A}{3} \left( \frac{k^2}{k_1^2} + \frac{\Delta_B(N_1)}{H^2(N_1)} \right) \right]^2\,,
\end{eqnarray}
where the power spectrum of the mode with $k < k_1$ at $N_1$ can be approximated by
$  \Delta_{\zeta_k}^2(N_1) \simeq  \Delta_{\epsilon}^2(N_k)$,
where $N_k$ is the time when the $k$-mode crosses out the horizon.
The dip is located at $k_{\rm dip}$ when  $|\Delta_{\zeta_k}(N_2)|^2$ reaches its minimum value, so we obtain
\begin{eqnarray}
k_{\rm dip} = k_1 \sqrt{ \frac{3}{A}-\frac{\Delta_B(N_1)}{H^2 (N_1)} }\,.
\end{eqnarray}
In the models under consideration, we have $\Delta_B(N_1)<0$ as a result of the loop contributions. Thus, the value of $k_{\rm dip}$ and its corresponding horizon cross-out $N$ will shift to a larger value as compared with that in the absence of loop corrections (i.e., $\Delta_B(N_1)=0$). This can be
seen in Fig.~\ref{figDeltaz} of Model~1. Such a shift is not so large for small values of $\Delta_B(N_1)$ as shown in Fig.~\ref{figDeltaz_2} of Model~2.
Therefore, the sign of $\Delta_B$ at the start of the USR phase may become important to locate the dip of the power spectrum in a general model.
Additionally, the large value of the amplification factor $A$ developed during the USR phase drives $k_{\rm dip}$ to a relatively
small value as seen in both Fig.~\ref{figDeltaz} and Fig.~\ref{figDeltaz_2},
and the steepest slope between $k_{\rm dip}$ and $k_1$ is about $k^4$.
Afterwards, the evolution of the power spectrum in the stage of the reentry into the SR phase does not significantly change the $k$ dependence.

For relatively large $k$ modes with $k_2 > k>k_1$ that cross out the horizon during the USR regime, the strong $k$ dependence of the power spectrum is apparent.
For such a $k$-mode which the effect of the loop correction (or the $\Delta_B$ term) can be ignored in its dynamical equation~(\ref{MS2a}), we  rewrite the equation in terms of the conformal time $\tau$ as
\begin{eqnarray}
\frac{d^2\zeta_k}{d\tau^2} + (2+\eta)aH\frac{d\zeta_k}{d\tau}+ k^2\zeta_k = 0. \label{zeta_large_k}
\end{eqnarray}
 The power spectrum for the $k$-mode at $N_2$ can be estimated by its evolution starting from $N_1$ to $N_k$ when the mode crosses out the horizon and then further to $N_2$ during the USR phase:
 \begin{equation}
 \Delta_{\zeta_k}^2 (N_2) \sim  \Delta_{\zeta_k}^2(N_1)   \left| e^{\int_{N_1}^{N_k}\left[ -(\eta+2)/2 + \sqrt{(\eta+2)^2 - 4k^2/(aH)^2}/2\right] dN} \right|^2  \left| \frac{k}{k_2} e^{\int_{N_k}^{N_2} \left[ -(\eta+2)/2 + |\eta+2|/2 \right] dN} \right|^2\,,
 \label{k1kk2}
 \end{equation}
where we have used approximate solutions of Eq.~(\ref{zeta_large_k}) for $\zeta_k$  and $\zeta_k'\equiv d\zeta_k/d\tau$, and $\eta \simeq -7$ which is almost a constant.
We have ignored the $k$ dependence in the evolution from $N_k$ to $N_2$ when $\zeta_k$ becomes a superhorizon mode with $k/a H <1$.
In addition, the evolution of the mode functions are such that
$|\zeta_k(N_k)| \sim |\dot{\zeta}_k(N_k)|/H < |\dot{\zeta}_k(N_2)|/H \sim |\zeta_k(N_2)|$, allowing us to estimate the evolution of $|\zeta_k|$ during $N_k$ and $N_2$ as
 \begin{equation}
\left| \frac{\zeta_k(N_2)}{\zeta_k(N_k)} \right| \sim \left| \frac{\dot{\zeta}_k(N_2)/H}{\dot{\zeta}_k(N_k)/H} \right| \sim \left| \frac{\zeta_k'(N_2)/a(N_2)H}{\zeta_k'(N_k)/a(N_k)H} \right| \sim \left| \frac{\zeta_k'(N_2)/k_2}{\zeta_k'(N_k)/k} \right| \,. \nonumber
 \end{equation}

In Eq.~(\ref{k1kk2}), the $k$ dependence in $\Delta_{\zeta_k}^2(N_1) $ can be extracted by approximating it as
 \begin{eqnarray}
 \Delta_{\zeta_k}^2(N_1)=\frac{k^3}{2\pi^2}  \left|\zeta_k(N_1)\right|^2 &\sim& \frac{k^3}{2\pi^2}  \left|\frac{1}{2a(N_1)\sqrt{k\epsilon(N_1)}}\right|^2 \nonumber\\
&\sim&  \Delta_\epsilon^2(N_1) \frac{k^2}{k_1^2}\,,
\end{eqnarray}
where we have used the asymptotic solution of Eq.~(\ref{zeta_large_k}),
\begin{equation}
\zeta_k \rightarrow \frac{1}{a\sqrt{2k}}\frac{e^{-ik\tau}}{\sqrt{2\epsilon}} \label{modefun_initial}
\end{equation}
as $-k\tau \gg 1$ for early times and evaluated it at $N_1$. At last, we arrive at
\begin{eqnarray}
\Delta_{\zeta_k}^2 (N_k) &\sim & \Delta_\epsilon^2(N_1)  \frac{k^2}{k_1^2}  \left| e^{\int_{N_1}^{N_k} -(\eta+2) dN} \right|  \left|  e^{ \int_{N_1}^{N_k} \sqrt{(\eta+2)^2 - 4k^2/(aH)^2} dN} \right|  \frac{k^2}{k_2^2} e^{-\int_{N_k}^{N_2} (2\eta + 4)dN}  \nonumber\\
&\sim & \Delta_\epsilon^2(N_1) \frac{k_1^{\eta}}{k_2^{2\eta+6}}k^{\eta+6}   \left|  e^{ \int_{N_1}^{N_k} \sqrt{(\eta+2)^2 - 4e^{-2N}k^2/H^2} dN} \right|\,.
\end{eqnarray}
With $\eta = -7$ during the USR regime, the power spectrum decreases as $k^{-1}$, resulting in a peak roughly near $N_1$ at the start of the USR regime. Including the evolution from $N_1$ to $N_k$ in the exponential factor shifts the peak location toward a larger $k$ or a larger $N$ that certainly depends on the details of a given model.

As for the $k$ modes that cross out the horizon after the USR phase with $k_2 < k < k_e$, where the $k_e$ mode just undergoes the horizon crossing at the end of inflation at $N_e$, their mode functions are such that  $|\zeta_k| > |\dot{\zeta}_k|/H$. The power spectrum at $N_e$ can be estimated by its evolution from $N_2$ till $N_e$ as
\begin{eqnarray}
\Delta_{\zeta_k}^2 (N_e)
&\sim &  \Delta_{\zeta_k}^2(N_2)   \left| e^{\int_{N_2}^{N_k}\frac{1}{2} \left[-(\eta+2)\pm\sqrt{ (\eta+2)^2-4k^2/(aH)^2} \right] dN} \right|^2   \left| e^{\int_{N_k}^{N_e} \left[-(\eta+2)+ (\eta+2) \right] dN} \right|  \nonumber \\
&\sim &  \Delta_\epsilon^2(N_2)  k_2^{\eta}k^{-\eta}  \left|  e^{ \int_{N_2}^{N_k} \sqrt{(\eta+2)^2 - 4e^{-2N}k^2/H^2} dN} \right| \nonumber \\
&\sim &
\Delta_\epsilon^2(N_k) \left|  e^{ \int_{N_2}^{N_k} \sqrt{(\eta+2)^2 - 4e^{-2N}k^2/H^2} dN} \right| \,,
\end{eqnarray}
where we have used the fact that
\begin{eqnarray}
 \Delta_{\zeta_k}^2(N_2)=\frac{k^3}{2\pi^2}  \left|\zeta_k(N_2)\right|^2 &\sim& \frac{k^3}{2\pi^2}  \left|\frac{1}{2a(N_2)\sqrt{k\epsilon(N_2)}}\right|^2 \nonumber\\
&\sim&  \Delta_\epsilon^2(N_2) \frac{k^2}{k_2^2}\,,
\end{eqnarray}
from the solution in Eq.~(\ref{modefun_initial}).   The slope of the power spectrum is about $k^{-\eta}$ and for $\eta >0$ in the second SR phase the power spectrum decreases
as $k$ increases. The slope of $\Delta_{\zeta_k}^2 (N)$ after $N_2$ shares the same slope of $\Delta_{\epsilon}^2 (N) \propto k^{-\eta}$ as shown in  Fig.~\ref{figDeltaz} and
Fig.~\ref{figDeltaz_2}.

\end{document}